\documentclass[conference]{IEEEtran}
\IEEEoverridecommandlockouts

\usepackage{graphicx}
\usepackage{url}
\usepackage[numbers,sort&compress]{natbib}
\usepackage[hidelinks]{hyperref}
\usepackage{float}

\usepackage{pdflscape}
\usepackage{array}
\usepackage{booktabs}
\usepackage{multirow}
\usepackage{caption}

\usepackage{xcolor}
\definecolor{thedarkblue}{RGB}{0,0,120}
\definecolor{mydarkblue}{rgb}{0,0.08,0.45}
\definecolor{darkblue}{rgb}{0,0.08,180}
\colorlet{TufteRed}{red!80!black}
\definecolor{theblue}{RGB}{0,0,180}
\colorlet{thered}{TufteRed}

\usepackage{microtype}
\usepackage{balance}
\usepackage{longtable}

\usepackage{booktabs}
\usepackage{multirow}
\usepackage{caption}
\usepackage{array}
\usepackage{pdflscape}
\usepackage{xcolor}

\usepackage{booktabs}
\usepackage{tabularx}

\usepackage{amsmath,amssymb,amsthm}

\usepackage{comment}

\usepackage{tikz}
\usepackage{verbatim}
\usetikzlibrary{arrows}
\usetikzlibrary{shapes,snakes}
\usetikzlibrary{decorations.pathmorphing}
\usetikzlibrary{fit}
\usetikzlibrary{backgrounds}

\usepackage{ragged2e}
\usepackage{multirow}
\usepackage{microtype}
\usepackage{balance}
\usepackage{setspace}

\graphicspath{{./}{./graphics/}}
\newcolumntype{H}{>{\setbox0=\hbox\bgroup}c<{\egroup}@{}}

\newcolumntype{R}[1]{>{\RaggedLeft\arraybackslash}}
\newcolumntype{L}[1]{>{\RaggedRight\arraybackslash}}

\AtBeginEnvironment{pmatrix}{\setlength{\arraycolsep}{2pt}}

\DeclareMathOperator{\hugeE}{\mbox{\huge\raise-0.3ex\hbox{E}}}
\DeclareMathOperator{\p}{\mathbb{P}}
\DeclareMathOperator{\hugep}{\mbox{\huge\raise-0.3ex\hbox{$\p$}}}

\usepackage[most]{tcolorbox}
\newtcolorbox{PromptTextBox}[1]{%
  colback=gray!5,
  colframe=black,
  boxrule=0.5pt,
  arc=2mm,
  width=\linewidth,
  title={#1},
  fonttitle=\bfseries,
  enhanced,
  left=2mm,
  right=2mm,
  top=1mm,
  bottom=1mm,
  boxsep=1mm
}

\makeatletter
\def\section{\@startsection{section}{1}{\z@}{1.2ex plus 0.3ex minus 0.3ex}%
{0.6ex plus 0.2ex minus 0ex}{\normalfont\normalsize\centering\scshape}}%
\def\subsection{\@startsection{subsection}{2}{\z@}{1.2ex plus 0.3ex minus 0.3ex}%
{0.6ex plus 0.2ex minus 0ex}{\normalfont\normalsize\itshape}}%
\makeatother
\AtBeginDocument{%
  \setlength{\abovedisplayskip}{4pt plus 1pt minus 1pt}%
  \setlength{\belowdisplayskip}{4pt plus 1pt minus 1pt}%
  \setlength{\abovedisplayshortskip}{2pt plus 1pt}%
  \setlength{\belowdisplayshortskip}{2pt plus 1pt}}

\renewcommand{\footnoterule}{\kern-3pt\hrule width 0.4\columnwidth\kern 2.6pt}
\begin{document}
\title{MusicRec: Audio, Lyrics, and Engagement Feature Enrichment for Music Recommendation}

\makeatletter
\newcommand{\linebreakand}{%
  \end{@IEEEauthorhalign}
  \hfill\mbox{}\par
  \mbox{}\hfill\begin{@IEEEauthorhalign}
}
\makeatother

\author{\IEEEauthorblockN{Srikar Prabhas Kandagatla\textsuperscript{1,*},
Sreehitha R. Narayana\textsuperscript{1,*},
Chandana Magapu\textsuperscript{1},
Swetha Mohan\textsuperscript{1},\\
Shamanth Kuthpadi Seethakantha\textsuperscript{1},
Hongjie Chen\textsuperscript{2},
Ryan A. Rossi\textsuperscript{3},
Franck Dernoncourt\textsuperscript{3} and
Nesreen Ahmed\textsuperscript{4}}
\IEEEauthorblockA{\textsuperscript{1}University of Massachusetts Amherst\\
Email: \{srikarprabhaskandagatla, sreehitha1707\}@gmail.com, \{hmagapu, swethamohan, skuthpadi\}@umass.edu}
\IEEEauthorblockA{\textsuperscript{2}Dolby Laboratories\\
Email: hongjie.chen@dolby.com}
\IEEEauthorblockA{\textsuperscript{3}Adobe Research\\
Email: \{ryrossi, dernonco\}@adobe.com}
\IEEEauthorblockA{\textsuperscript{4}Cisco Research\\
Email: nesahmed@cisco.com}}

\maketitle
\begingroup
\renewcommand{\thefootnote}{}%
\footnotetext{\fontsize{8.5pt}{11pt}\selectfont $^{*}$Equal contribution.}%
\footnotetext{\fontsize{8.5pt}{11pt}\selectfont This paper has been accepted at the 2026 IEEE International Conference on Big Data (IEEE BigData 2026). This version includes additional experiments and appendices. \copyright~2026 IEEE. Personal use of this material is permitted. Permission from IEEE must be obtained for all other uses, in any current or future media, including reprinting/republishing this material for advertising or promotional purposes, creating new collective works, for resale or redistribution to servers or lists, or reuse of any copyrighted component of this work in other works.}%
\endgroup
\pagestyle{plain}
\thispagestyle{plain}

\begin{abstract}
Music recommendation systems typically treat songs as opaque tokens, relying on interaction histories that overlook semantic or acoustic content.
Prior LLM-augmented, multimodal, and text-enhanced sequential recommenders partially combine semantic, acoustic, or engagement signals, and none jointly model all three within a unified LLM-based framework grounded in song content.
In this work, we propose a multimodal framework for session-based music recommendation that enriches the LastFM-1K dataset with three complementary signals: (1) audio and lyric embeddings from pretrained music and text models, (2) LLM-generated semantic metadata using the MGPHot schema, and (3) listening completion ratios.
We validate the generated metadata against human MGPHot annotations across all 58 attributes.
We extend the E4SRec framework with multimodal features, ItemID encoder backbones including SASRec, BERT4Rec, and GRU4Rec, and LLM backbones including LLaMA-2-13B, Qwen2.5-7B-Instruct, and LLaMA-3-70B across zero-shot and fine-tuned settings.
In addition, we evaluate four multimodal fusion strategies, benchmark five audio and five lyric encoders, and compare Spotify audio attributes with LLM-generated MGPHot metadata.
Our experiments show that content-based features improve over ID-only baselines by up to 95\% in Recall and 79\% in NDCG, with the largest gains when the ItemID baseline is weakest. Fine-tuning substantially improves over zero-shot, after which the best audio or lyric setup outperforms the ID-only baseline for every backbone.
Naive multimodal fusion does not always yield additive improvements, highlighting challenges in cross-modal integration, while structured LLM-generated metadata proves more useful than Spotify audio attributes.
We release a large-scale multimodal benchmark for music recommendation research.
\end{abstract}

\newpage
\section{Introduction}
Sequential music recommendation aims to predict the next song a user is likely to listen to based on their listening history~\cite{tran2024pisa, hansen2020contextual}. Modern recommender systems use architectures where songs are represented as discrete item identifiers and user behavior is modeled as sequences of item interactions~\cite{li2023e4srec,hansen2020contextual}.
While these ID-based approaches perform well in dense interaction settings, they treat each track as an independent token and fail to capture its underlying content signals.
As a result, they struggle under sparsity and cold-start conditions and cannot effectively reason about acoustic similarity, lyrical themes, or semantic context~\cite{li2025mdsbr,xu2025mdvt, giahi2025vlclip}.
Music preference is inherently multimodal, influenced by acoustic properties such as tempo and timbre, lyrical content, and higher-level semantic attributes including mood, instrumentation, and genre~\cite{moscati2022music4all}.
Recent studies suggest that incorporating multimodal information can produce richer item representations and improve recommendation quality~\cite{vaswani2021multimodal,oramas2017deep}.
For example, TalkPlay~\cite{talkplay} demonstrates that grounding LLM-based music recommenders with audio, lyrics, and semantic metadata improves both recommendation relevance and conversational quality. Despite these advances, multimodal grounding has not been systematically explored within sequential music recommendation frameworks.
Additionally, although Large Language Models (LLMs) have been used for recommendation for
their ability to model sequential behavior and reason over textual
inputs~\cite{liao2024llara},
most existing LLM-based recommenders rely heavily on ID-based embeddings or weak textual proxies, limiting their grounding in the actual content of songs~\cite{li2023e4srec,zheng2024adaptinglargelanguagemodels}.

Intuitively, enriching song representations with multimodal information
may address these limitations.
In this work, we bridge this gap by constructing a multimodal music
recommendation framework on top of the \textit{LastFM-1K} dataset~\cite{lastfm1k}. We
enrich song representations with three complementary modalities: (1) audio and lyric embeddings representing acoustic and semantic content of the song, (2) LLM-generated semantic metadata describing musical characteristics such as harmony, rhythm, sonority, instrumentation, and lyrical themes, and (3) behavioral engagement signals derived from listening completion ratios.
These enriched representations are integrated to a recommendation framework \textit{E4SRec}~\cite{li2023e4srec} to systematically evaluate how different modalities
contribute to recommendation performance.
Our contributions are as follows:
\begin{itemize}
    \item We propose a pipeline for curating multimodal song semantic data, incorporating audio and lyric embeddings, LLM-generated semantic metadata, and song completion-ratio signals.
    The pipeline is validated with experiments using music recommendation framework E4SRec.

    \item We conduct comprehensive experiments by extending E4SRec with multiple sequential recommendation backbones, including \textit{SASRec, BERT4Rec,} and \textit{GRU4Rec}, and multiple LLM backbones, including \textit{LLaMA-3-70B, LLaMA-2-13B,} and \textit{Qwen2.5-7B-Instruct}. We further evaluate four fusion strategies under both zero-shot and fine-tuned settings across all backbones.

    \item We release a multimodal music recommendation benchmark built from LastFM-1K, supporting future studies on integrating multimodal features into LLM-based recommendation systems. The dataset is openly available at:\\\url{https://doi.org/10.5281/zenodo.20431748}.
\end{itemize}

\section{Related Work}
We discuss related work in four categories:
sequential recommendation, session-based recommendation, multimodal recommendation, and music representations.

\medskip\noindent\textbf{\textit{Sequential Recommendation.}}
By modeling the chronological order of past interactions, recommenders in this setting capture how user preferences evolve over time. Recent advances in Transformers, Graph Neural Networks, and LLM-based recommenders have significantly improved their ability to capture long-range dependencies and multimodal user behavior~\cite{liu2025ccfrec}. Self-attention models such as SASRec~\cite{kang2018sasrec} and BERT4Rec~\cite{sun2019bert4rec} encode item sequences, and we use both as ItemID backbones.
\citet{li2023e4srec} proposed \textit{E4SRec}, which integrates LLMs with ID-based sequential recommenders by directly injecting pretrained ItemID embeddings into the input layer and replacing autoregressive generation with a single-pass prediction layer. Using LoRA~\cite{hu2022lora} for efficient adaptation, E4SRec achieves state-of-the-art accuracy and mitigates data sparsity without rich semantic descriptions.
\citet{wang2025intent} proposed \textit{IRLLRec}, a model-agnostic framework that bridges the semantic gap between unstructured text and structured interactions via a dual-tower architecture with pairwise and translation alignment strategies, and outperforms collaborative filtering baselines.
\citet{tran2024pisa} proposed \textit{PISA}, a Transformer-based sequential recommender tailored for music listening that integrates ACT-R cognitive memory mechanisms to capture repetitive listening behavior. On Last.fm and Deezer, PISA predicts engagement with both new and repeated tracks.
\citet{du2025not} proposed \textit{IntervalLLM}, which incorporates irregular time intervals into LLMs via an Interval-Infused Attention mechanism, outperforming baselines by an average of 4.4\% across overall, warm, and cold-start scenarios.

\begin{figure*}[t]
\centering
\includegraphics[width=0.92\textwidth]{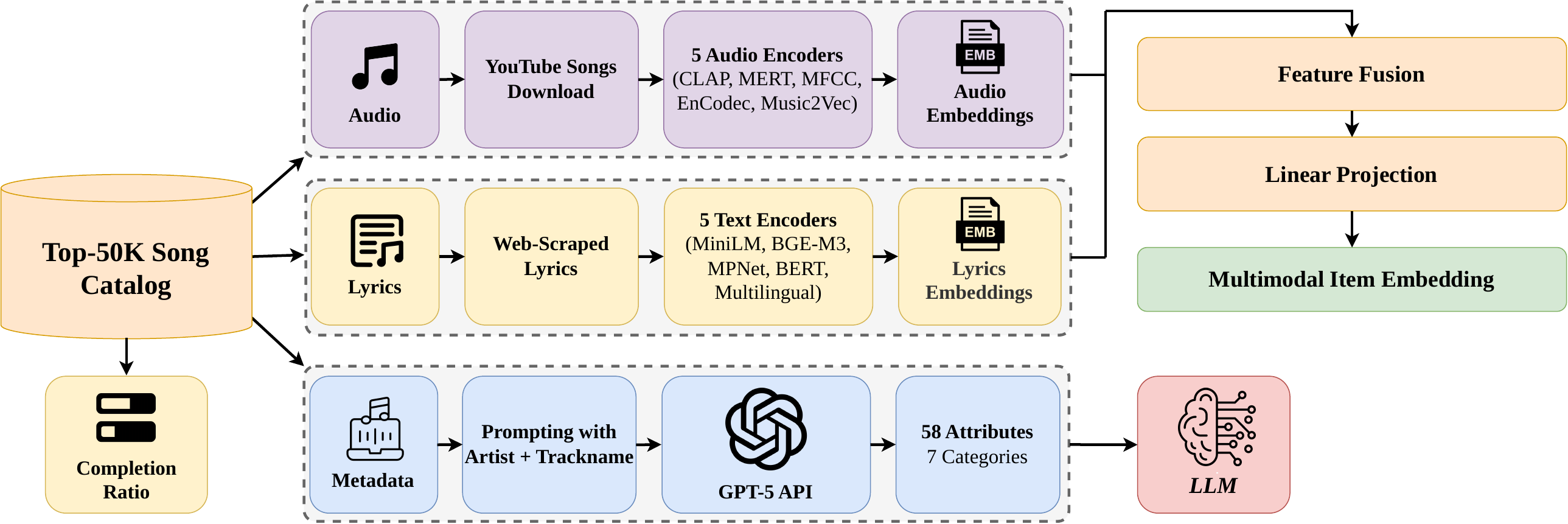}
\caption{Multimodal Song Enrichment Pipeline: Audio, Lyric, Metadata, and Engagement Signal Extraction.
}
\label{fig:feature-extraction}
\end{figure*}

\medskip\noindent\textbf{\textit{Session-based Recommendation.}}
Next-item prediction from only the interactions within a short user session, without long-term user profiles, is particularly important in music streaming, where user intent shifts rapidly and contextual listening behavior strongly influences track selection. Early neural approaches encode the session with a recurrent network, as in GRU4Rec~\cite{hidasi2016gru4rec}.
\citet{yu2020tagnn} proposed \textit{TAGNN}, a graph-based session recommendation model that represents each session as a directed graph and introduces a target-aware attention mechanism that computes candidate-specific session representations. By dynamically attending to different parts of the session depending on the target item, TAGNN achieves more expressive next-item prediction, particularly relevant for music sessions containing multiple micro-intents.
\citet{hansen2020contextual} study large-scale music recommendation in real-world streaming systems, learning dynamic user representations that evolve based on recent listening activity. Accounting for temporal dynamics significantly improves their next-track prediction at production scale.
While these works effectively capture behavioral signals from interaction sequences, neither incorporates richer semantic representations beyond collaborative signals. Our approach extends session-based modeling by integrating LLM-generated metadata and audio embeddings, enabling deeper understanding of song characteristics within evolving listening sessions of individual users.

\medskip\noindent\textbf{\textit{Multimodal Recommendation.}}
Adding heterogeneous content signals such as text, audio, images, and structured metadata to interaction data extends traditional recommender systems beyond collaborative signals.
\citet{qin2024atflrecmultimodalrecommenderaudiotext} proposed \textit{ATFLRec}, an audio-text fusion framework built on an instruction-tuned LLM that jointly models cross-modal interactions within the LLM space using LoRA for efficient adaptation, outperforming unimodal and graph-based baselines. Unlike ATFLRec, which assumes access to raw audio, we use externally derived audio embeddings, enabling scalable multimodal modeling without raw waveforms.
\citet{lyu2025xreflectcrossreflectionpromptingmultimodal} proposed \textit{X-Reflect}, a cross-reflection prompting framework that explicitly prompts MLLMs to reconcile supportive and conflicting signals between text and images, producing more nuanced item representations and outperforming existing prompting baselines.
\citet{pomo2025recommendersystemsreallyleverage} show that naive fusion strategies often fail to achieve meaningful cross-modal alignment, and that Large Vision-Language Models with structured prompting produce semantically coherent embeddings that outperform simple feature concatenation, directly motivating our use of LLM-generated metadata alongside audio embeddings.
TalkPlay~\cite{talkplay} grounds conversational music recommendation on audio, lyrics, and auto-captioned tags, and targets multi-turn dialogue rather than session-based sequential prediction.
Music4All-Onion~\cite{moscati2022music4all} combines audio, tag, and lyric features over Last.fm, without engagement signals such as completion ratio.
Overall, effective cross-modal alignment is critical for recommendation robustness, yet most existing approaches focus on static item enrichment and assume full modality access. We address both by combining audio embeddings with LLM-generated metadata.

\medskip\noindent\textbf{\textit{Representation for Music Recommendation.}}
Accurate user and item representations are the cornerstone of modern music recommendation systems. By mapping sparse interaction signals into dense vectors, a recommender can measure similarity, compute relevance scores, or feed generative models.

The most traditional approach learns a low-dimensional vector per user and per track directly from the interaction matrix. Methods such as BPR~\cite{rendle2009bpr}, LightGCN~\cite{he2020lightgcn}, and NGCF~\cite{wang2019ngcf} capture pure collaborative signals and yield strong ranking results on dense datasets, but performance degrades sharply under data sparsity or cold-start conditions due to tight coupling with observed interactions.
To alleviate sparsity, many works augment or replace ItemID vectors with dense acoustic representations extracted from the raw waveform or spectrogram, using back-ends such as \textit{MusiCNN}~\cite{van2019musicscnn}, \textit{MERT}~\cite{li2023mert}, \textit{Music2Vec}~\cite{chen2022music2vec}, \textit{MusicFM}~\cite{won2023musicfm}, and \textit{EnCodecMAE}~\cite{pepino2023encodecmae}.
These item-side embeddings are either frozen or fine-tuned jointly with the recommendation loss, and empirically improve Recall and NDCG on sparse playlists~\cite{tang2024pretrainedaudio,yuan2023morec}, especially when combined with collaborative ID vectors.
When metadata such as titles, lyrics, tags, or cover art is available, BERT/RoBERTa-style models and ResNet/ViT encoders can supply rich item-side vectors, while systems like CLAP~\cite{wu2023large} align audio and text in a shared space for cross-modal retrieval.
Knowledge graph (KG) methods such as TransE, RGCN, and CompGCN model item attributes and relations alongside collaborative vectors. KDR~\cite{mu2021kdr} aligns these KG embeddings with collaborative factors via mutual-information maximization, yielding interpretable, disentangled factors such as mood, instrumentation, and tempo. Matryoshka Representation Learning~\cite{zhang2022matryoshka} instead builds a hierarchy of increasingly fine-grained vectors. Both approaches mitigate over-generalization and improve diversity-aware and long-tail recommendation metrics.

In summary, representation learning for music recommendation has progressed from pure ID-based collaborative vectors to rich, multimodal, and semantically structured embeddings, with a direct and measurable effect on recommendation quality, especially under data sparsity. We combine audio and lyric embeddings, LLM-generated metadata, and completion ratios in one session-aware LLM recommender.

\section{Multimodal Data Curation Pipeline}
We propose a multimodal data curation pipeline for music recommendation that enriches the public LastFM-1K dataset (Sec.~\ref{sec:prelim}) beyond interaction-only item representations. Each track is augmented with audio and lyric embeddings (Sec.~\ref{sec:audio-lyric-embeds}), LLM-generated metadata (Sec.~\ref{sec:prelim} and~\ref{sec:LLM-gen-features}), and completion-ratio signals (Sec.~\ref{sec:cr}), then fused as shown in Fig.~\ref{fig:feature-extraction}. The curated data enables evaluating multimodal features in zero-shot and fine-tuned sequential recommendation.

\subsection{Preliminary}
\label{sec:prelim}

\medskip\noindent\textbf{\textit{Dataset.}}
We build our benchmark on top of the \textit{LastFM-1K} dataset~\cite{lastfm1k}, a public collection of timestamped listening histories from about 1,000 Last.fm users. Each record contains a user ID, timestamp, artist, track, and MusicBrainz IDs.
To construct a session-based recommendation dataset, we filter users with fewer than 1,000 listening events and tracks with fewer than 7 total plays. Following prior work on session-based music recommendation~\cite{tran2024pisa, hansen2020contextual}, we segment user histories into sessions by splitting interactions separated by more than 20 minutes of inactivity, and sessions with fewer than 10 interactions are discarded.
The resulting dataset contains 814 users, 4.2 million listening events, and 295,957 unique tracks spanning February 2005 to June 2009. Table~\ref{tab:lastfm-stats} summarizes the processed dataset.
The benchmark is diverse in both language and genre. Approximately 27.1\% of songs are non-English tracks spanning 42 language codes, while genre annotations cover 81 unique genres. These properties make the dataset suitable for studying recommendation robustness across heterogeneous and multilingual musical content.
We further construct a Top-50k song catalog by selecting the 50,029 most frequently played tracks from the filtered interaction data.

\medskip\noindent\textbf{\textit{Musicological Features from MGPHot}.}
To capture structured musicological properties, we annotate songs using the
\textit{MGPHot} schema~\cite{oramas2025mgphot}, which defines 58 musicological
attributes spanning lyrics, vocals, harmony, rhythm, instrumentation,
sonority, and composition. These attributes provide interpretable
descriptions of musical characteristics not observable from interaction histories alone. Table~\ref{tab:mgphot-cats} lists the seven categories and their representative attributes used in this work.

\begin{table}[t]
\centering
\small
\caption{Preprocessed LastFM-1K dataset statistics.}
\label{tab:lastfm-stats}
\resizebox{\linewidth}{!}{

\begin{tabular}{lrlr}
\toprule
\textbf{Metric} & \textbf{Value} & \textbf{Metric} & \textbf{Value} \\
\midrule
Users & 814
& Total scrobbles & 4.21M \\

Unique tracks & 295{,}957
& Unique artists & 43{,}406 \\

Sessions & 421{,}396
& Sparsity & 0.0175 \\

Mean session duration & 44.39 min
& Median session duration & 41.33 min \\
\midrule
\multicolumn{1}{l}{Date range}
& \multicolumn{2}{l}{2005-02-14 to 2009-06-19} \\
\bottomrule
\end{tabular}
}
\end{table}
\begin{table}[t]
\centering
\small
\setlength{\tabcolsep}{4pt}
\caption{MGPHot attribute categories and features.}
\label{tab:mgphot-cats}
\resizebox{\linewidth}{!}{
\begin{tabular}{ll}
\toprule
\textbf{Category} & \textbf{Representative attributes} \\
\midrule
Lyrics          & Angry, Sad, Happy, Humorous, Love (mood dimensions) \\
Vocals          & Register, Timbre (Thin to Full), Breathiness, Smoothness \\
Harmony         & Minor/Major Key Tonality, Harmonic Sophistication \\
Rhythm          & Tempo, Swing Feel, Syncopation, Danceability \\
Instrumentation & Drum Set, Electric Guitar, Acoustic Guitar, Piano \\
Sonority        & Live Recording, Acoustic, Electric \\
Composition     & Focus on Lead Vocal, Focus on Melody, Focus on Lyrics \\
\bottomrule
\end{tabular}
}
\end{table}

\begin{table}[t]
\centering
\small

{
\renewcommand{\arraystretch}{0.85}
\caption{Extended musicological features curated from a three-model rating consensus of LLaMA-3.3-70B-Instruct, Qwen2.5-7B-Instruct, and Mistral-Nemo-12B-Instruct.
}
\label{tab:llm-metafeatures}
\resizebox{0.8\linewidth}{!}{

\begin{tabular}{lll}
\toprule
\textbf{Cat.} & \textbf{Attribute} & \textbf{Values} \\
\midrule

\multirow{9}{*}{\rotatebox[origin=c]{90}{Lyric structure}}
& \multirow{3}{*}{Rhyme scheme} & AABB / ABAB \\
& & ABCB / mixed \\
& & irregular / free \\
& Internal rhyme density & [0, 1] \\
& Rhyme complexity score & 0--4 \\
& Repetition ratio & [0, 1] \\
& Vocabulary richness & 1--5 \\
& Profanity intensity & 0--4 \\
& Rhyme-pattern note & free text \\
\midrule

\multirow{3}{*}{\rotatebox[origin=c]{90}{Narrative}}
& \multirow{2}{*}{Dominant perspective} & 1st / 2nd \\
& & 3rd / mixed \\
& Perspective shift & yes / no \\
& Shift description & free text \\
\midrule

\multirow{6}{*}{\rotatebox[origin=c]{90}{Novelty}}
& Title-content coherence & [0, 1] \\
& Genre subversion score & 0--4 \\
& Humor / irony presence & 0--2 \\
& Spoken-word vs.\ sung ratio & [0, 1] \\
& Humor type & free text \\
& Subversion rationale & free text \\

\bottomrule
\end{tabular}
}
}
\end{table}
\medskip\noindent\textbf{\textit{Extended Musicological Features.}}
In addition to the MGPHot schema, we extract a broader set of song-level meta-features using prompted LLM inference over lyrics and track metadata. These features target meaning, structure, and cultural context not captured by audio encoders or by metadata fields such as genre tags or Spotify audio attributes. Table~\ref{tab:llm-metafeatures} summarizes the three categories: lyric structure, narrative, and novelty.

\subsection{Audio and Lyric Embeddings}
\label{sec:audio-lyric-embeds}
\medskip\noindent\textbf{\textit{Audio.}}
We augment the song catalog with two sets of audio-derived features: handcrafted acoustic descriptors and dense neural audio embeddings. These representations complement semantic metadata by capturing sonic characteristics that cannot be inferred directly from lyrics or textual descriptions alone.

For the handcrafted descriptors, we compute a 63-dimensional feature vector using Librosa~\cite{mcfee2015librosa} for each song, comprising: 40 Mel-Frequency Cepstral Coefficients (MFCCs, capturing timbre), 2 spectral centroid values (brightness), 14 spectral contrast values (harmonic texture), 2 spectral rolloff values (high-frequency energy distribution), 2 RMS energy values (loudness), 2 zero-crossing rate values (noise versus tonal signal distinction), and 1 tempo estimate (beats per minute). Additionally, structured audio attributes such as valence, energy, and danceability are retrieved for 47{,}113 songs (94.2\% coverage) via the ReccoBeats API.

For the richer acoustic embeddings, we encode each song using four pre-trained audio encoders: \textit{CLAP}~\cite{wu2023large} (music-language contrastive pretraining), \textit{MERT}~\cite{li2023mert} (music-specific self-supervised pretraining), \textit{Music2Vec}~\cite{chen2022music2vec} (contrastive self-supervised learning on music audio), and \textit{EnCodec}~\cite{defossez2022encodec} (neural audio codec representation). In addition to these neural encoders, we include MFCC-based representations as a classical handcrafted baseline.
Audio files are retrieved through automated matching and download pipelines using track and artist metadata from the LastFM-1K catalog.

\medskip\noindent\textbf{\textit{Lyrics.}} We generate lyric embeddings with five pretrained text encoders, \textit{MiniLM}, \textit{BGE-M3}, \textit{MPNet}, \textit{Multilingual Sentence Encoder (MultiLG)}, and \textit{BERT}~\cite{wang2020minilm,chen2024bgem3,song2020mpnet,reimers2020multilingual,devlin2019bert}, using lyrics from LRCLIB and Genius, with Whisper~\cite{radford2023whisper} as a fallback.

Audio-derived features complement semantic metadata by modeling timbre, tempo, energy, instrumentation, and production characteristics that are not explicitly represented in text-based descriptions alone. The high feature coverage across the catalog enables large-scale multimodal recommendation experiments under both zero-shot and fine-tuned settings.

Some of these attributes deliberately complement rather than duplicate existing signals. For instance, the profanity intensity feature provides a graded measure of explicit content, going beyond Spotify's binary explicit flag. Temporal orientation (past/present/future framework) is orthogonal to valence and captures nostalgia or aspiration independently of positivity/negativity. The spoken word ratio distinguishes rap and poetry from sung melody in a way that is qualitatively different from Spotify's \textit{speechiness}, which conflates the two. Together, these meta-features extend the semantic reach of the benchmark by exposing lyric-level and cultural dimensions of songs. These properties are difficult to infer reliably from either audio embeddings or structured acoustic descriptors alone.

\begin{figure}[t]
\centering
\ttfamily\scriptsize
\begin{PromptTextBox}{MGPHot Feature Rating Prompt Template}
\textcolor{black!85}{\textbf{SYSTEM:}}
Rate the song on each MGPHot musical attribute as an integer 0-5:\\
0 - absent/lowest;\\
5 - dominant/highest;\\
Return only a JSON object with the requested keys. No explanation.\\

\textcolor{black!85}{\textbf{USER:}}
"Mrs. Robinson" by Simon \& Garfunkel\\
Keys: ["Vocal Register", "Vocal Timbre Thin to Full", "Focus on Performance", etc.]\\

\textcolor{black!85}{\textbf{ASSISTANT:}}
\{"Vocal Register": 4, "Vocal Timbre Thin to Full": 4, "Vocal Breathiness": 4, "Focus on Performance": 4, etc.\}\\

\textcolor{black!85}{\textbf{USER:}}
"Master Of Puppets" by Metallica\\
Keys: ["Vocal Register", "Vocal Timbre Thin to Full", "Focus on Performance", etc.]\\

\textcolor{black!85}{\textbf{ASSISTANT:}}
\{"Vocal Register": 4, "Vocal Timbre Thin to Full": 4, "Vocal Breathiness": 0, "Focus on Performance": 4, etc.\}\\

\textcolor{black!85}{\textbf{USER:}}
"\textit{\{title\}}" by \textit{\{artist\}}\\
Keys: ["Vocal Register", "Vocal Timbre Thin to Full", "Focus on Performance", etc.]
\end{PromptTextBox}

\caption{Prompt template for MGPHot feature rating.}

\label{fig:mgphot-prompt}
\end{figure}

\subsection{LLM-Generated Musicological Features}
\label{sec:LLM-gen-features}

\medskip\noindent\textbf{\textit{Musicological Features from MGPHot.}} For each track, we query an LLM, Azure OpenAI GPT-5~\cite{openai2025gpt5}, via the Batch API, with the artist name and track title and ask it to rate every MGPHot
attribute on an Likert-type scale from 0 (absent / lowest) to 5
(dominant / highest). The full prompt template is shown in Fig.~\ref{fig:mgphot-prompt}. After validating on a held-out subset of the original MGPHot annotations, we selected a \emph{few-shot} prompting variant that prepends two multi-turn anchor examples whose ground-truth ratings are taken directly from the published MGPHot annotations: one predominantly acoustic and vocal track \emph{Mrs.\ Robinson} by Simon \& Garfunkel and one heavy and electric track, \emph{Master of Puppets} by Metallica. These two anchors are chosen to satisfy three criteria simultaneously. First, they occupy opposite extremes across the primary MGPHot dimensions, including instrumentation profile, vocal character, energy,
and production style, which provides the model with maximally separated reference points from which to interpolate ratings for unseen tracks. Second, both tracks are sufficiently canonical that the underlying LLM possesses dense, reliable pretraining knowledge of their sonic properties, reducing the risk of hallucinated attribute values. Third, their ground-truth MGPHot ratings are unambiguous and span the full 0 to 5 scale on complementary attribute subsets, ensuring that the few-shot context exercises the entire rating range rather than clustering near one end. Together, these anchors calibrate the model's use of the 0 to 5 range across stylistic extremes. Raw integer ratings are subsequently rescaled to $[0, 1]$ by dividing by 5, yielding continuous-valued feature vectors that describe properties such as vocal timbre, harmonic sophistication, rhythmic complexity, instrumentation profile, emotional tone, and production style.

\medskip\noindent\textbf{\textit{Annotation Reliability.}}
We validate the LLM ratings against the MGPHot human annotations~\cite{oramas2025mgphot} on 2,195 held-out tracks covered by both sources, reaching an overall Spearman $\rho$ of 0.45 and MAE of 0.14 across all 58 attributes (Appendix~\ref{app:mgphot-validation}, Table~\ref{tab:MGPHot-validation} and Fig.~\ref{fig:mgphot-eval-summary}).
Agreement is strongest for lyrics, sonority, and instrumentation, easiest to infer from artist and title, and weakest for composition and rare attributes.

\medskip\noindent\textbf{\textit{Extended Musicological Features.}}
For each track, we also utilize an LLM to extract the extended musicological features via zero-shot prompting on the artist name, track title, and available lyric text, returning structured values for each feature.

\subsection{Song Completion Ratio}
\label{sec:cr}
On the interaction side, we add a completion ratio signal that relates listening duration to track length, estimated from the gap to the next play (Sec.~\ref{sec:cr-signal}). A user who listens to nearly an entire song expresses a different preference than one who skips after a few seconds, yet conventional recommenders treat both interactions identically. Completion ratios let the framework distinguish strong engagement from incidental exposure.

\section{Recommendation Method}

We formulate session-based music recommendation as a next-song prediction problem.
Given a user listening session $s_u = (i_1, i_2, \ldots, i_T)$, where each $i_t \in \mathcal{I}$ denotes a song from the candidate catalog, the goal is to predict the next song $i_{T+1}$ by ranking all candidate songs.
For each song $i$, we construct a multimodal representation by fusing four inputs,
\[
\mathbf{z}_i =
\phi\left(
\mathbf{e}^{\mathrm{id}}_i,
\mathbf{a}_i,
\mathbf{l}_i,
\mathbf{m}_i
\right),
\]
where $\mathbf{e}^{\mathrm{id}}_i$ is the ItemID embedding, $\mathbf{a}_i$ is the audio embedding, $\mathbf{l}_i$ is the lyric embedding, $\mathbf{m}_i$ is the metadata feature embedding, and $\phi(\cdot)$ is a multimodal fusion function.
For each interaction in the session, we incorporate the completion ratio $c_{u,i_t} \in [0,1]$ to form an engagement-aware item representation:
\[
\mathbf{x}_{u,i_t} = g(\mathbf{z}_{i}, c_{u,i_t}).
\]

The session encoder $R$ maps the sequence of engagement-aware item representations to a user-state vector:
\[
\mathbf{h}_u^t = R(\mathbf{x}_{u,i_1}, \mathbf{x}_{u,i_2}, \ldots, \mathbf{x}_{u,i_t}).
\]

The next song is predicted by scoring each candidate item and selecting the highest-ranked candidate song:
\[
\begin{aligned}
\hat{i}_{T+1}
&=
\arg\max_{i \in \mathcal{I}}
f\left(\mathbf{h}_u^t, \mathbf{z}_i\right) \\
&=
\arg\max_{i \in \mathcal{I}}
f\left(
R(\mathbf{x}_{u,i_1}, \ldots, \mathbf{x}_{u,i_t}),
\mathbf{z}_i
\right).
\end{aligned}
\]

\begin{figure}[t]
\centering
\includegraphics[width=0.92\linewidth]{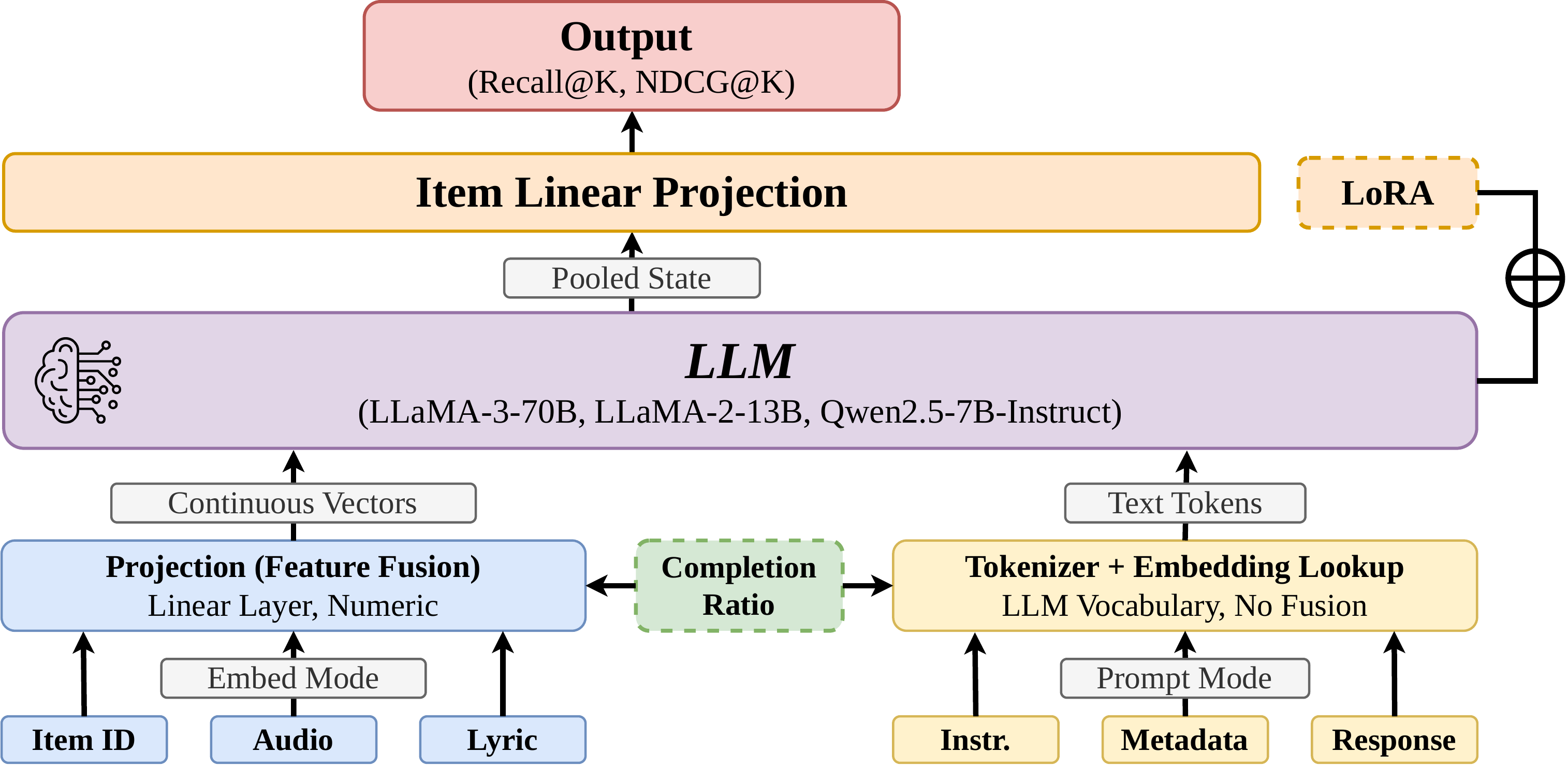}
\caption{Proposed multimodal recommendation architecture.
}
\label{fig:model_architecture}
\end{figure}

\begin{figure*}[t]
\centering
\includegraphics[width=0.8\textwidth]{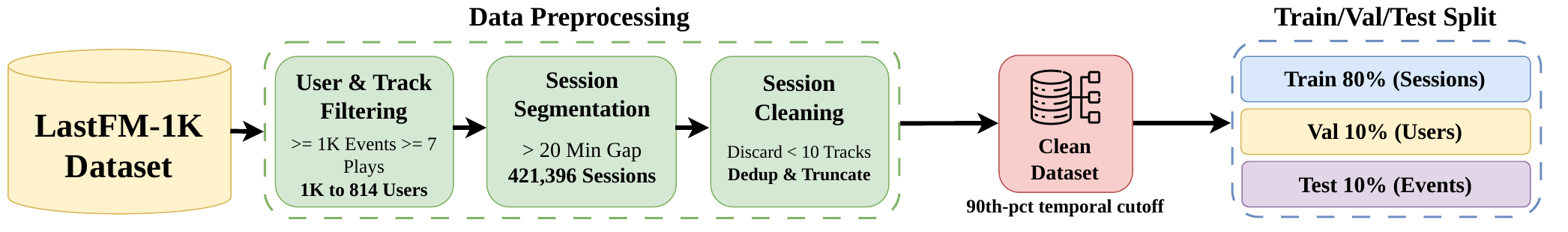}
\caption{Data preprocessing and splits.}
\label{fig:data-preprocessing}
\end{figure*}

We enable the base recommender $R$ to process multimodal item representations and engagement-aware interaction signals. Content features from audio, lyrics, and structured metadata enrich item representations beyond ID-based embeddings, capturing similarity between songs at the level of sound, lyrical meaning, and musical attributes. The completion ratio further serves as an engagement signal, letting the model distinguish strong listening behavior from incidental exposure.

As illustrated in Fig.~\ref{fig:model_architecture}, ItemIDs, multimodal features, and completion ratios are provided with instruction prompts, allowing the LLM-based recommender $R$ to condition predictions on both item content and user engagement.

\subsection{Multimodal Item Representations}
\label{sec:multimodal-item-repre}

For each song $i$, we combine the ItemID embedding with the audio, lyric, and metadata features of Sec.~\ref{sec:audio-lyric-embeds} and~\ref{sec:LLM-gen-features}. Audio captures acoustic properties, lyrics encode semantic and emotional content, and metadata describes high-level musical attributes. The recommender can then compare songs by both collaborative patterns and content similarity.

\begin{table}[t]
\centering
\caption{Top 50k song catalog statistics.}
\label{tab:catalog-stats}
\small
\setlength{\tabcolsep}{4pt}
\resizebox{\linewidth}{!}{%
\begin{tabular}{llll}
\toprule
\multicolumn{2}{c}{\textbf{General}} &
\multicolumn{2}{c}{\textbf{Coverage / Distribution}} \\
\cmidrule(r){1-2} \cmidrule(l){3-4}
\textbf{Metric} & \textbf{Value} &
\textbf{Metric} & \textbf{Value} \\
\midrule
Total songs              & 50{,}029 &
Audio features           & 47{,}113 (94.2\%) \\

Unique artists           & 5{,}751 &
Preview URL              & 32{,}142 (64.2\%) \\

Avg.\ play count         & 49.7 &
Genre tag                & 28{,}105 (56.2\%) \\

Median play count        & 31.0 &
Unique genres            & 81 \\

Avg.\ duration (min)     & 4.14 &
English                  & 36{,}373 (72.7\%) \\

Top genre: Rock          & 10{,}774 &
Non-English              & 13{,}536 (27.1\%) \\

Electronic               & 3{,}463 &
Unknown language         & 120 (0.2\%) \\

Pop                      & 1{,}441 &
Unique language codes    & 42 \\
\bottomrule
\end{tabular}%
}
\end{table}
\subsection{Item Representation and LLM Injection}
\label{sec:llm-injection}
Each ItemID backbone is first trained to convergence with next-item prediction, yielding frozen 64-dimensional embeddings $\mathbf{e}^{\mathrm{id}}_i$ over the Top-50k catalog (Table~\ref{tab:catalog-stats}). Audio, lyric, and metadata embeddings are precomputed offline (Sec.~\ref{sec:audio-lyric-embeds}) and stored as frozen matrices indexed by the same ItemID.

The fusion function $\phi$ of Sec.~\ref{sec:fusion-strategies} combines the available modality vectors into $\mathbf{z}_i$. When the completion ratio is injected in embed mode, the scalar is appended before projection,
\[
g(\mathbf{z}_i, c_{u,i_t}) = [\,\mathbf{z}_i \,\Vert\, c_{u,i_t}\,],
\]
which widens the projection input by one dimension. In prompt mode, $g$ is the identity and $c_{u,i_t}$ is rendered as text on the corresponding history line. A single linear layer $W_p$ maps each engagement-aware item vector into the LLM hidden size,
\[
\mathbf{u}_{u,i_t} = W_p \, g(\mathbf{z}_i, c_{u,i_t}),
\]
and the resulting sequence is placed between the tokenized instruction and the tokenized response marker,
\[
X_u = [\,E(\mathrm{instr}) \,\Vert\, \mathbf{u}_{u,i_1}, \ldots, \mathbf{u}_{u,i_t} \,\Vert\, E(\mathrm{resp})\,],
\]
where $E(\cdot)$ denotes the LLM token embedding. Semantic metadata enters through $E(\mathrm{instr})$ as natural language, while ID, audio, and lyric information enters as projected vectors. Fig.~\ref{fig:model_architecture} shows both paths, one in embed mode with continuous vectors and one in prompt mode with text tokens.

\medskip\noindent\textbf{\textit{Training and Inference.}}
The final hidden state of $X_u$ gives the user state $\mathbf{h}_u^t$, and a bias-free linear head $W_s$ scores items. LoRA~\cite{hu2022lora} adapts the LLM, and the fusion module, $W_p$, and $W_s$ are trained jointly with all embedding tables frozen. Training minimizes a sampled softmax cross-entropy over the ground-truth item $i^{+}$ and 128 negatives $\mathcal{N}$ resampled per step, keeping memory independent of catalog size:
\[
\mathcal{L} = -\log \frac{\exp\!\left(\mathbf{h}_u^t \cdot W_s[i^{+}]\right)}{\exp\!\left(\mathbf{h}_u^t \cdot W_s[i^{+}]\right) + \sum_{j \in \mathcal{N}} \exp\!\left(\mathbf{h}_u^t \cdot W_s[j]\right)}.
\]
At inference, the same head scores the evaluation candidate set, and items are ranked by $\mathbf{h}_u^t \cdot W_s[i]$.

\subsection{Engagement-Aware Signaling}
\label{sec:cr-signal}

In standard sequential recommendation, each observed interaction is usually treated as a positive event. However, music listening behavior contains additional implicit feedback. A user who listens to nearly the entire track likely expresses a stronger preference than a user who skips the track after a short duration, yet both events receive the same positive label.

To capture this distinction, we incorporate completion ratio as an engagement-aware signal. For a user interaction with song $i$, the completion ratio is defined as follows:
\[
c_i = \min\!\left(1, \frac{d_i^{o}}{d_i^{tot}}\right), \qquad d_i^{o} = \max\left(0,\, t_{i+1} - t_i\right),
\]
where $d_i^{o}$ is the time to the next interaction in the same session, used as the observed listening duration because LastFM-1K logs no playback duration, and $d_i^{tot}$ is the total duration of the song. The ratio is undefined for the last interaction of a session. Higher values of $c_i$ indicate stronger engagement, while lower values may correspond to skips or weak preference.

We use this signal to provide the model with a more informative view of user behavior within a session. Rather than assuming that all historical items contribute equally to the next-item prediction task, completion ratio allows the framework to distinguish strongly preferred songs from incidental or incomplete listening events such as early skips.

\subsection{Multimodal Fusion Strategies}
\label{sec:fusion-strategies}

We evaluate four fusion strategies to study how different modalities should be combined: (1) Concatenation combines ItemID, audio, and lyric embeddings along the feature dimension followed by a linear projection, (2) Weighted sum learns scalar weights for each modality and combines them into a single representation, (3) Cross-attention treats the ItemID embedding as a query and the modality embeddings as keys and values, allowing the model to selectively attend to relevant content features, (4) FiLM~\cite{perez2018film} applies feature-wise affine modulation, where content embeddings generate scaling and shifting parameters that condition the ItemID representation.

Each strategy is applied per modality, fusing the ItemID embedding with the audio embedding, the lyric embedding, or both using the same strategy. When the completion ratio is added in prompt or embed mode (Sec.~\ref{sec:llm-injection}), the audio and lyric embeddings are combined by concatenation.

These fusion strategies provide different trade-offs between simplicity, parameter efficiency, and cross-modal expressiveness. By evaluating them under the same recommendation framework, we analyze whether multimodal information is most useful when directly combined, selectively attended to, or used to modulate the collaborative item representation.

\section{Experiments} \label{sec:exp}

\begin{table*}[t]
\centering
\small
\setlength{\tabcolsep}{4pt}
\caption{Zero-shot (LLaMA-3-70B) and fine-tuned (LLaMA-2-13B) results across sequential encoders and input configurations.}
\label{tab:experiment-main-results-combined}
\resizebox{\textwidth}{!}{
\begin{tabular}{lccccccccccccccccccc}
\toprule
& & & & & & \multicolumn{6}{c}{\textbf{Zero-shot: LLaMA-3-70B}}
& \multicolumn{6}{c}{\textbf{Fine-tuned: LLaMA-2-13B}} \\
\cmidrule(lr){7-12} \cmidrule(lr){13-18}
\multirow{2}{*}{\textbf{Seq. Model}}
& \multirow{2}{*}{\textbf{ID}}
& \multirow{2}{*}{\textbf{Audio}}
& \multirow{2}{*}{\textbf{Lyric}}
& \multirow{2}{*}{\textbf{Meta}}
& \multirow{2}{*}{\textbf{CR}}
& \multicolumn{2}{c}{@5} & \multicolumn{2}{c}{@10} & \multicolumn{2}{c}{@20}
& \multicolumn{2}{c}{@5} & \multicolumn{2}{c}{@10} & \multicolumn{2}{c}{@20} \\
\cmidrule(lr){7-8} \cmidrule(lr){9-10} \cmidrule(lr){11-12}
\cmidrule(lr){13-14} \cmidrule(lr){15-16} \cmidrule(lr){17-18}
& & & & & & Recall & NDCG & Recall & NDCG & Recall & NDCG
& Recall & NDCG & Recall & NDCG & Recall & NDCG \\
\midrule
\multirow{3}{*}{SASRec}
& \checkmark &            &            &            &
& 0.022 & 0.015 & 0.049 & 0.023 & 0.100 & 0.036
& 0.229 & 0.125 & \textbf{0.323} & 0.155 & 0.398 & 0.174 \\
& \checkmark & \checkmark & \checkmark & \checkmark &
& \textbf{0.024} & \textbf{0.016} & \textbf{0.053} & \textbf{0.025} & \textbf{0.119} & \textbf{0.042}
& 0.234 & 0.127 & 0.317 & 0.154 & \textbf{0.404} & \textbf{0.177} \\
& \checkmark & \checkmark & \checkmark & \checkmark & \checkmark
& 0.020 & 0.015 & 0.049 & 0.024 & 0.091 & 0.034
& \textbf{0.236} & \textbf{0.128} & 0.316 & \textbf{0.156} & 0.402 & 0.175 \\
\midrule
\multirow{3}{*}{BERT4Rec}
& \checkmark &            &            &            &
& 0.020 & 0.014 & 0.042 & 0.021 & 0.089 & 0.033
& 0.224 & 0.122 & 0.321 & 0.154 & 0.392 & 0.172 \\
& \checkmark & \checkmark & \checkmark & \checkmark &
& 0.022 & 0.013 & 0.047 & 0.021 & 0.107 & 0.036
& 0.232 & 0.127 & 0.318 & 0.155 & 0.393 & 0.174 \\
& \checkmark & \checkmark & \checkmark & \checkmark & \checkmark
& \textbf{0.039} & \textbf{0.025} & \textbf{0.063} & \textbf{0.033} & \textbf{0.124} & \textbf{0.048}
& \textbf{0.237} & \textbf{0.129} & \textbf{0.323} & \textbf{0.156} & \textbf{0.400} & \textbf{0.176} \\
\midrule
\multirow{3}{*}{GRU4Rec}
& \checkmark &            &            &            &
& 0.025 & 0.015 & 0.044 & 0.021 & 0.088 & 0.032
& 0.228 & 0.124 & 0.316 & 0.152 & 0.393 & 0.172 \\
& \checkmark & \checkmark & \checkmark & \checkmark &
& \textbf{0.033} & \textbf{0.020} & \textbf{0.056} & \textbf{0.028} & 0.103 & 0.039
& \textbf{0.236} & \textbf{0.129} & \textbf{0.327} & \textbf{0.158} & \textbf{0.396} & \textbf{0.176} \\
& \checkmark & \checkmark & \checkmark & \checkmark & \checkmark
& 0.025 & 0.018 & 0.049 & 0.025 & \textbf{0.114} & \textbf{0.042}
& 0.231 & 0.126 & 0.322 & 0.156 & 0.390 & 0.173 \\
\bottomrule
\end{tabular}
}
\end{table*}

We evaluate zero-shot and fine-tuned recommendation, and then study the impact of encoders, metadata, and fusion.

\subsection{Experimental Setup} \label{sec:exp-setup}

\medskip\noindent\textbf{\textit{Dataset Preprocessing and Data Splits.}}
We preprocess the LastFM-1K dataset, as described in Sec.~\ref{sec:prelim}, for our experiments.
The statistics are provided in Table~\ref{tab:catalog-stats}.
The LastFM-1K dataset provides no off-the-shelf splits. We first segment listening histories into sessions following \citet{tran2024pisa}, then define train/validation/test splits using the global temporal cutoff methodology of \citet{abbattista2024}. The 90th-percentile timestamp serves as the test cutoff, placing the most
recent 10\% of interactions in the test set. For validation, $\sim$10\% of
users ($\approx$81, sampled from the 814-user cohort) are randomly selected and an analogous 90th-percentile temporal cutoff is applied within their pre-test history; 60 of these end up with a non-empty validation window. The final train / val / test cohorts contain 638 / 60 / 638 users respectively. The same fixed splits are applied to all models to ensure fair comparison, with the validation set used exclusively for hyperparameter tuning and the test set is reserved solely for final evaluation and never used during training or development.
Fig.~\ref{fig:data-preprocessing} shows the full preprocessing and data split pipeline, from raw LastFM-1K listening events to the final train, validation, and test cohorts.

\medskip\noindent\textbf{\textit{Baseline.}}
We adopt E4SRec~\cite{li2023e4srec}, a recent LLM-based sequential recommendation method and the closest baseline to our setting. E4SRec maps pretrained ItemID embeddings into the LLM input space, enabling the language model to leverage user interaction sequences for next-item prediction.

This vanilla E4SRec is used as baseline, which is especially relevant because it captures sequential user behavior through item-level representations but does not explicitly incorporate multimodal song information such as audio, lyrics, metadata, or engagement-aware signals. Therefore, it provides a controlled reference point for assessing the contribution of our multimodal item representations and fusion strategies. For consistency, we re-train and evaluate E4SRec on the same preprocessed LastFM-1K splits and report results using the same evaluation protocol and ranking metrics as all our multimodal variants.

\medskip\noindent\textbf{\textit{Evaluation Metrics.}}
We evaluate all model variants on a fixed held-out test set across four experimental settings: (1) ID-only baselines, (2) semantic metadata augmentation, (3) content embedding augmentation with audio and lyric embeddings, and (4) engagement-aware modeling. This staged evaluation setup enables controlled analysis of the contribution of each enrichment component to recommendation performance.

We evaluate recommendation quality using standard top-k ranking metrics computed over the test set. For each test session, the model ranks all candidate items and the held-out ground-truth item is evaluated within the ranked candidate list.

We adopt Recall@$K$ and Normalized Discounted Cumulative Gain (NDCG@$K$) at $K \in \{5, 10, 20\}$ as our primary metrics, consistent with the evaluation practice of our baselines, which enables direct comparison. Recall@$K$ measures whether the ground-truth item appears within the top-$K$ recommendations, rewarding any correct retrieval regardless of rank. NDCG@k (Normalized Discounted Cumulative Gain) is a position-sensitive metric that assigns higher scores to correct items ranked closer to the top of the list, providing a more nuanced measure of ranking quality. The encoder comparison additionally reports MRR and Precision at $K \in \{5, 10, 20\}$, and the metadata study reports MRR@10.

\medskip\noindent\textbf{\textit{Zero-shot versus Finetuning Settings.}}
In zero-shot experiments, LLM weights are frozen and no task-specific training is performed. Item representations are constructed from the multimodal features described in Sec.~\ref{sec:multimodal-item-repre} and Sec.~\ref{sec:cr-signal}, combined using the fusion strategies described in Sec.~\ref{sec:fusion-strategies} and projected into the LLM hidden space. In contrast, fine-tuning experiments use LoRA-based parameter-efficient adaptation applied to the LLM backbone. Fine-tuning experiments use the same backbone \& modality configuration grid as zero-shot experiments to enable direct comparison.

\medskip\noindent\textbf{\textit{Implementation Details.}}
Zero-shot and fine-tuned runs share the same backbone and modality grid. Fine-tuning applies LoRA~\cite{hu2022lora} (r=32, alpha=64, dropout=0.05) to all attention and MLP projection layers of the 4-bit-quantized backbone, optimized with paged AdamW-8bit under a cosine schedule at peak learning rate 1e-4, warmup ratio 0.1, batch size 64, and 5 epochs. Checkpoints are selected by lowest loss on a held-out set of 500 interactions. Zero-shot inference keeps every parameter frozen, with LoRA adapters held at initialization. The metadata prompt carries attributes whose normalized rating exceeds 0.4, item history covers the 10 most recent interactions, and the numeric ID sequence retains up to 256. LLaMA-3-70B~\cite{grattafiori2024llama3} and LLaMA-2-7B~\cite{touvron2023llama2} run zero-shot only, while LLaMA-2-13B and Qwen2.5-7B-Instruct~\cite{qwen2024qwen25} run in both.

\subsection{Zero-Shot Results}
Table~\ref{tab:experiment-main-results-combined} (left) shows the results from the E4SRec framework with the LLaMA-3-70B backbone with various ItemID encoder backbones. The primary finding is that multimodal enrichment continues to improve recommendation quality in several zero-shot settings, but the gains are highly dependent on the combination of the sequential encoder and input configuration. Additionally, the results show that the effect of completion ratio is particularly architecture-dependent. For BERT4Rec, adding audio, lyric, metadata, and completion ratio produces the best performance across all metrics, improving Recall@20 from 0.089 to 0.124 and NDCG@20 from 0.033 to 0.048. This represents the strongest LLaMA-3-70B configuration and suggests that completion ratio provides a useful behavioral signal when combined with BERT4Rec representations. In contrast, SASRec achieves its best performance with audio, lyric, and metadata features alone, reaching Recall@20 of 0.119 and NDCG@20 of 0.042, while adding completion ratio reduces Recall@20 to 0.091 and NDCG@20 to 0.034. GRU4Rec shows a cutoff-dependent trend, where audio, lyric, and metadata improve Recall@5 and NDCG@10, but the completion-ratio variant performs better for K=20.

\subsection{Fine-Tuned Results}

Table~\ref{tab:experiment-main-results-combined} (Right) presents the fine-tuned results using the LLaMA-2-13B. For SASRec, the ID-only baseline already attains a strong Recall@10 of 0.323. The Audio + Lyric + Metadata configuration achieves the best Recall@20 of 0.404 and NDCG@20 of 0.177, while adding completion ratio slightly reduces performance, reaching Recall@10 of 0.316 and Recall@20 of 0.402 but yields marginal gains at shorter cut-offs, achieving Recall@5 of 0.236 from 0.229 and NDCG@5 of 0.128 from 0.125, indicating a cutoff-dependent sensitivity to completion ratio under fine-tuning. For BERT4Rec, the full configuration, including audio, lyric, metadata, and completion ratio, consistently achieves the best performance across all metrics, improving Recall@5 from 0.224 to 0.237, Recall@10 from 0.321 to 0.323, and Recall@20 from 0.392 to 0.400. This suggests that completion ratio provides a reliable behavioral signal for BERT4Rec even in the fine-tuned setting. For GRU4Rec, the Audio + Lyric + Metadata configuration yields the strongest overall results, with NDCG@10 of 0.158 and Recall@10 of 0.327, while the inclusion of the completion ratio does not yield further gains, suggesting limited incremental benefit for GRU4Rec under fine-tuning.

\subsection{Experiments with Additional LLMs}
\label{sec:additional-backbones}
We further evaluate two additional LLM backbones, LLaMA-2-13B and Qwen2.5-7B-Instruct, under zero-shot inference.
As shown in Table~\ref{tab:experiment-ablation-study-across-LLMs-zero-shot-small}, adding audio, lyric, and metadata features often improves both Recall@K and NDCG@K over the interaction-only setting, suggesting that content-based signals provide useful complementary information beyond ItemID sequence representations in the zero-shot setting.

Under LLaMA-2-13B, SASRec benefits most clearly from multimodal augmentation. Adding audio, lyric, and metadata features improves Recall@10 from 0.061 to 0.078, a relative gain of 27.9\%, and improves NDCG@10 from 0.026 to 0.035, a relative gain of 34.6\%. For BERT4Rec and GRU4Rec, the completion-ratio variant gives slight improvements, such as NDCG@20 from 0.038 to 0.039 for BERT4Rec and Recall@10 from 0.061 to 0.063 for GRU4Rec, while the gains are less consistent than in the SASRec setting.

For Qwen2.5-7B-Instruct, the strongest overall performance is obtained with GRU4Rec using the interaction-only configuration, achieving Recall@20 of 0.108 and NDCG@20 of 0.038. However, multimodal enrichment is still beneficial for BERT4Rec. In particular, BERT4Rec improves from 0.102 to 0.121 in Recall@20. This indicates that multimodal features are especially useful when the sequential encoder does not already capture sufficiently strong item-transition patterns.

\begin{table}[t]
\centering
\scriptsize
\setlength{\tabcolsep}{0.8pt}
\caption{Additional LLM backbones: zero-shot results using our benchmark across LLM backbones and configurations.}
\label{tab:experiment-ablation-study-across-LLMs-zero-shot-small}
\resizebox{\linewidth}{!}{
\begin{tabular}{c l l ccccc cccccc@{}}
\toprule
\multirow{2}{*}{\rotatebox{90}{LLM}}
& \multirow{2}{*}{Model}
& \multicolumn{5}{c}{Configuration}
& \multicolumn{2}{c}{@5}
& \multicolumn{2}{c}{@10}
& \multicolumn{2}{c}{@20} \\
\cmidrule(lr){3-7}
\cmidrule(lr){8-9}
\cmidrule(lr){10-11}
\cmidrule(lr){12-13}
& & ID & Audio & Lyric & Meta & CR
& Recall & NDCG
& Recall & NDCG
& Recall & NDCG \\
\midrule

\multirow{9}{*}{\rotatebox{90}{LLaMA-2-13B}}
& \multirow{3}{*}{SASRec}
  & \checkmark &  &  &  &
  & 0.030 & 0.016 & 0.061 & 0.026 & \textbf{0.127} & 0.042 \\
& & \checkmark & \checkmark & \checkmark & \checkmark &
  & \textbf{0.038} & \textbf{0.022} & \textbf{0.078} & \textbf{0.035} & 0.116 & \textbf{0.045} \\
& & \checkmark & \checkmark & \checkmark & \checkmark & \checkmark
  & 0.013 & 0.007 & 0.036 & 0.014 & 0.088 & 0.027 \\
\cmidrule(lr){2-13}

& \multirow{3}{*}{BERT4Rec}
  & \checkmark &  &  &  &
  & 0.024 & 0.014 & 0.056 & 0.025 & \textbf{0.108} & 0.038 \\
& & \checkmark & \checkmark & \checkmark & \checkmark &
  & 0.024 & 0.015 & 0.045 & 0.023 & 0.091 & 0.034 \\
& & \checkmark & \checkmark & \checkmark & \checkmark & \checkmark
  & \textbf{0.033} & \textbf{0.019} & \textbf{0.058} & \textbf{0.028} & 0.102 & \textbf{0.039} \\
\cmidrule(lr){2-13}

& \multirow{3}{*}{GRU4Rec}
  & \checkmark &  &  &  &
  & 0.030 & \textbf{0.019} & 0.061 & 0.029 & \textbf{0.125} & \textbf{0.045} \\
& & \checkmark & \checkmark & \checkmark & \checkmark &
  & \textbf{0.031} & 0.016 & 0.049 & 0.022 & 0.099 & 0.034 \\
& & \checkmark & \checkmark & \checkmark & \checkmark & \checkmark
  & \textbf{0.031} & \textbf{0.019} & \textbf{0.063} & \textbf{0.030} & 0.107 & 0.041 \\
\midrule

\multirow{9}{*}{\rotatebox{90}{Qwen2.5-7B-Instruct}}
& \multirow{3}{*}{SASRec}
  & \checkmark &  &  &  &
  & \textbf{0.033} & \textbf{0.020} & \textbf{0.056} & \textbf{0.027} & \textbf{0.100} & \textbf{0.038} \\
& & \checkmark & \checkmark & \checkmark & \checkmark &
  & 0.024 & 0.014 & 0.045 & 0.021 & 0.097 & 0.034 \\
& & \checkmark & \checkmark & \checkmark & \checkmark & \checkmark
  & 0.016 & 0.010 & 0.039 & 0.017 & 0.088 & 0.029 \\
\cmidrule(lr){2-13}

& \multirow{3}{*}{BERT4Rec}
  & \checkmark &  &  &  &
  & \textbf{0.033} & \textbf{0.020} & 0.055 & 0.026 & 0.102 & 0.038 \\
& & \checkmark & \checkmark & \checkmark & \checkmark &
  & 0.031 & 0.017 & \textbf{0.063} & \textbf{0.027} & \textbf{0.121} & \textbf{0.041} \\
& & \checkmark & \checkmark & \checkmark & \checkmark & \checkmark
  & 0.031 & 0.019 & 0.055 & 0.026 & 0.102 & 0.038 \\
\cmidrule(lr){2-13}

& \multirow{3}{*}{GRU4Rec}
  & \checkmark &  &  &  &
  & \textbf{0.025} & \textbf{0.015} & \textbf{0.058} & \textbf{0.026} & \textbf{0.108} & \textbf{0.038} \\
& & \checkmark & \checkmark & \checkmark & \checkmark &
  & 0.022 & 0.013 & 0.044 & 0.020 & 0.097 & 0.033 \\
& & \checkmark & \checkmark & \checkmark & \checkmark & \checkmark
  & 0.017 & 0.010 & 0.034 & 0.015 & 0.083 & 0.027 \\

\bottomrule
\end{tabular}
}
\end{table}

\begin{table}[t]
\centering
\scriptsize
\setlength{\tabcolsep}{0.8pt}
\caption{Audio and lyric embedding performance comparison across encoders. SASRec is the ID-only baseline. Bold indicates the best result per encoder group.}
\label{tab:audio-lyric-encoder-results}
\resizebox{\linewidth}{!}{
\begin{tabular}{c l cccc c l cccc}
\toprule
& \multicolumn{5}{c}{\textbf{Audio Embeddings}}
&
& \multicolumn{5}{c}{\textbf{Lyric Embeddings}} \\
\cmidrule(lr){2-6} \cmidrule(lr){8-12}
\textbf{K}
& \textbf{Model} & \textbf{Recall} & \textbf{NDCG} & \textbf{MRR} & \textbf{Prec.}
&
& \textbf{Model} & \textbf{Recall} & \textbf{NDCG} & \textbf{MRR} & \textbf{Prec.} \\
\midrule

\multirow{6}{*}{5}
& SASRec    & 0.024 & 0.014 & 0.011 & 0.005 & & SASRec  & 0.024 & 0.014 & 0.011 & 0.005 \\
& CLAP      & 0.022 & 0.015 & 0.012 & 0.004 & & MiniLM  & \textbf{0.050} & 0.019 & 0.010 & \textbf{0.010} \\
& MERT      & 0.033 & 0.022 & 0.019 & \textbf{0.007} & & BGE-M3  & 0.000 & 0.000 & 0.000 & 0.000 \\
& Music2Vec & 0.025 & 0.015 & 0.012 & 0.005 & & MPNet   & \textbf{0.050} & \textbf{0.050} & \textbf{0.050} & \textbf{0.010} \\
& EnCodec   & \textbf{0.034} & \textbf{0.024} & \textbf{0.020} & \textbf{0.007} & & MultiLG & \textbf{0.050} & 0.032 & 0.025 & \textbf{0.010} \\
& MFCC      & 0.017 & 0.008 & 0.005 & 0.003 & & BERT    & 0.000 & 0.000 & 0.000 & 0.000 \\

\midrule

\multirow{6}{*}{10}
& SASRec    & 0.039 & 0.019 & 0.013 & 0.004 & & SASRec  & 0.039 & 0.019 & 0.013 & 0.004 \\
& CLAP      & 0.039 & 0.020 & 0.014 & 0.004 & & MiniLM  & \textbf{0.100} & 0.035 & 0.016 & \textbf{0.010} \\
& MERT      & 0.058 & 0.030 & 0.022 & 0.006 & & BGE-M3  & 0.000 & 0.000 & 0.000 & 0.000 \\
& Music2Vec & 0.044 & 0.021 & 0.014 & 0.004 & & MPNet   & 0.050 & \textbf{0.050} & \textbf{0.050} & 0.005 \\
& EnCodec   & \textbf{0.072} & \textbf{0.035} & \textbf{0.025} & \textbf{0.007} & & MultiLG & 0.050 & 0.032 & 0.025 & 0.005 \\
& MFCC      & 0.047 & 0.018 & 0.009 & 0.005 & & BERT    & 0.000 & 0.000 & 0.000 & 0.000 \\

\midrule

\multirow{6}{*}{20}
& SASRec    & 0.089 & 0.032 & 0.016 & 0.004 & & SASRec  & 0.089 & 0.032 & 0.016 & 0.004 \\
& CLAP      & 0.085 & 0.031 & 0.017 & 0.004 & & MiniLM  & 0.100 & 0.035 & 0.016 & 0.005 \\
& MERT      & 0.108 & 0.042 & 0.025 & 0.005 & & BGE-M3  & \textbf{0.150} & 0.037 & 0.010 & \textbf{0.008} \\
& Music2Vec & 0.086 & 0.032 & 0.017 & 0.004 & & MPNet   & 0.100 & \textbf{0.061} & \textbf{0.053} & 0.005 \\
& EnCodec   & \textbf{0.121} & \textbf{0.048} & \textbf{0.028} & \textbf{0.006} & & MultiLG & 0.050 & 0.032 & 0.025 & 0.002 \\
& MFCC      & 0.086 & 0.027 & 0.012 & 0.004 & & BERT    & 0.050 & 0.011 & 0.002 & 0.002 \\

\bottomrule
\end{tabular}
}
\end{table}

\subsection{Impact of Encoders}

\medskip\noindent\textbf{\textit{Audio Embeddings.}}
 We evaluate the usefulness of acoustic embeddings over five audio embedders CLAP~\cite{wu2023large}, MERT~\cite{li2023mert}, Music2Vec~\cite{chen2022music2vec}, EnCodec~\cite{defossez2022encodec} and MFCC independently under zero-shot inference with Qwen2.5-7B-Instruct. Table~\ref{tab:audio-lyric-encoder-results} (Left) reports Recall, NDCG, MRR, and Precision at $K \in \{5,10,20\}$.
EnCodec achieves the strongest overall performance across most metrics and cutoffs, consistently outperforming the SASRec baseline, so we select EnCodec as the primary audio embedder for the main experiments. The spread is wide, with CLAP at 0.085 and MFCC at 0.086 below the SASRec row at Recall@20 (0.089), indicating that a weak acoustic representation degrades ranking rather than improving it. MERT ranks second with Recall@20 of 0.108, suggesting that music-specific pretraining captures more recommendation-relevant acoustic detail than general-purpose CLAP.

\medskip\noindent\textit{\textbf{Lyric Embeddings.}}
Similar to audio embedders, we evaluate all five lyric embedders (MiniLM, BGE-M3, MPNet, MultiLG, and BERT) independently under zero-shot inference with Qwen2.5-7B-Instruct. Table~\ref{tab:audio-lyric-encoder-results} (Right) reports Recall, NDCG, MRR, and Precision at k $\in$ \{5, 10, 20\}.
Among the five lyric embedders, BERT and BGE-M3 fail to produce meaningful recommendations at tight cutoffs ($k \leq 10$), scoring near zero across all metrics, suggesting weaker alignment between these representations and the recommendation objective. MPNet achieves the best NDCG and MRR across all cutoffs, despite not always achieving the highest Recall. MiniLM achieves high Recall at k=10 but lower ranking precision, and BGE-M3 improves at larger cutoffs primarily through broader retrieval coverage rather than precise ranking. We select MPNet as the primary lyric embedder for its consistently superior ranking quality.
Recall@20 spans 0.050 to 0.150 on lyrics and 0.085 to 0.121 on audio, with each gap between best and worst encoders exceeding the modality's gain over SASRec. Encoder choice matters more than modality, one likely reason multimodal gains are inconsistent across backbones and settings.

\begin{table}[t]
\centering
\small
\setlength{\tabcolsep}{0.8pt}
\caption{Zero-shot performance across (LLM, ItemID backbone, metadata) configurations, reported in Recall@20, NDCG@10, and MRR@10. The best results are highlighted.
}
\label{tab:zero-shot-all}
\resizebox{\linewidth}{!}{
\begin{tabular}{lcccccccccc}
\toprule
\multirow{2}{*}{\textbf{Config}} & \multirow{2}{*}{CR} & \multicolumn{3}{c}{\textbf{LLaMA-3-70B}} & \multicolumn{3}{c}{\textbf{LLaMA-2-7B}} & \multicolumn{3}{c}{\textbf{Qwen2.5-7B-Inst.}} \\
\cmidrule(lr){3-5} \cmidrule(lr){6-8} \cmidrule(lr){9-11}
&  & SAS. & BERT. & GRU. & SAS. & BERT. & GRU. & SAS. & BERT. & GRU. \\
\midrule
\multicolumn{11}{c}{\textit{Recall@20}} \\
\midrule
Spotify &            & 0.085          & 0.089          & 0.099          & 0.089          & 0.092          & 0.077          & \textbf{0.103} & 0.086          & 0.094 \\
Spotify & \checkmark & 0.097          & 0.085          & 0.099          & \textbf{0.111} & 0.107          & 0.096          & 0.088          & 0.092          & 0.102 \\
MGPHot  &            & 0.097          & 0.088          & 0.092          & 0.085          & 0.099          & 0.099          & 0.121          & \textbf{0.122} & 0.089 \\
MGPHot  & \checkmark & 0.099          & 0.097          & 0.091          & 0.096          & 0.100          & \textbf{0.119} & 0.078          & 0.114          & 0.091 \\
S+M     &            & 0.089          & 0.097          & 0.100          & \textbf{0.107} & 0.100          & 0.086          & 0.086          & 0.085          & 0.096 \\
S+M     & \checkmark & 0.096          & 0.102          & \textbf{0.114} & 0.097          & 0.074          & 0.099          & 0.080          & 0.091          & 0.103 \\
\midrule
\multicolumn{11}{c}{\textit{NDCG@10}} \\
\midrule
Spotify &            & 0.018          & 0.013          & 0.021          & 0.019          & 0.020          & 0.019          & \textbf{0.022} & 0.016          & 0.015 \\
Spotify & \checkmark & 0.019          & 0.014          & 0.019          & \textbf{0.023} & 0.019          & 0.019          & 0.017          & 0.015          & 0.019 \\
MGPHot  &            & \textbf{0.024} & 0.014          & 0.015          & 0.013          & 0.018          & 0.018          & 0.023          & 0.021          & 0.016 \\
MGPHot  & \checkmark & 0.018          & 0.015          & 0.013          & 0.016          & 0.019          & \textbf{0.021} & 0.013          & 0.020          & 0.016 \\
S+M     &            & 0.016          & 0.017          & 0.014          & \textbf{0.019} & 0.016          & 0.014          & 0.010          & 0.014          & 0.018 \\
S+M     & \checkmark & 0.017          & \textbf{0.022} & 0.019          & 0.014          & 0.012          & 0.016          & 0.016          & 0.018          & 0.016 \\
\midrule
\multicolumn{11}{c}{\textit{MRR@10}} \\
\midrule
Spotify &            & 0.010          & 0.007          & 0.011          & 0.011          & 0.011          & 0.012          & \textbf{0.013} & 0.009          & 0.008 \\
Spotify & \checkmark & 0.010          & 0.007          & 0.009          & \textbf{0.013} & 0.010          & 0.011          & 0.009          & 0.008          & 0.011 \\
MGPHot  &            & 0.012          & 0.008          & 0.008          & 0.008          & 0.010          & 0.010          & \textbf{0.013} & 0.012          & 0.010 \\
MGPHot  & \checkmark & 0.010          & 0.008          & 0.007          & 0.008          & 0.010          & \textbf{0.012} & 0.007          & 0.011          & 0.008 \\
S+M     &            & 0.008          & 0.009          & 0.008          & \textbf{0.011} & 0.008          & 0.008          & 0.005          & 0.008          & 0.010 \\
S+M     & \checkmark & 0.008          & \textbf{0.013} & 0.012          & 0.008          & 0.006          & 0.009          & 0.009          & 0.010          & 0.009 \\
\bottomrule
\end{tabular}
}
\end{table}

\begin{table*}[t]
\centering
\scriptsize
\setlength{\tabcolsep}{3pt}
\caption{Zero-shot results for LLaMA-3-70B.}
\label{tab:LLaMa-3-70B-zeroshot-results}
\resizebox{0.83\textwidth}{!}{
\begin{tabular}{cllcccccccccc}
\toprule
\multirow{2}{*}{LLM}
& \multirow{2}{*}{Seq. Model}
& \multirow{2}{*}{ID}
& \multirow{2}{*}{Audio}
& \multirow{2}{*}{Lyric}
& \multirow{2}{*}{CR}
& \multicolumn{2}{c}{@5}
& \multicolumn{2}{c}{@10}
& \multicolumn{2}{c}{@20} \\
\cmidrule(lr){7-8} \cmidrule(lr){9-10} \cmidrule(lr){11-12}
& & & & & & Recall & NDCG & Recall & NDCG & Recall & NDCG \\
\midrule

\multirow{51}{*}{\rotatebox[origin=c]{90}{LLaMA-3-70B}}
& \multirow{17}{*}{SASRec}
& \checkmark & -- & -- & -- & 0.022 & 0.015 & 0.049 & 0.023 & 0.100 & 0.036 \\
& & \checkmark & -- & -- & Embed & 0.020 & 0.009 & 0.038 & 0.015 & 0.075 & 0.024 \\
& & \checkmark & -- & -- & Prompt & 0.025 & 0.019 & 0.052 & 0.028 & 0.100 & 0.040 \\
& & \checkmark & Concat & -- & -- & 0.027 & 0.013 & 0.053 & 0.021 & 0.086 & 0.029 \\
& & \checkmark & Weighted Avg. & -- & -- & 0.022 & 0.012 & 0.050 & 0.022 & 0.113 & 0.037 \\
& & \checkmark & Cross-Attn. & -- & -- & 0.028 & 0.018 & \textbf{0.063} & 0.029 & \textbf{0.116} & 0.042 \\
& & \checkmark & FiLM & -- & -- & \textbf{0.030} & \textbf{0.022} & 0.061 & \textbf{0.032} & 0.113 & \textbf{0.045} \\
\cmidrule(r){3-12}
& & \checkmark & -- & Concat & -- & 0.020 & 0.013 & \textbf{0.049} & 0.022 & 0.091 & 0.032 \\
& & \checkmark & -- & Weighted Avg. & -- & 0.020 & 0.011 & 0.033 & 0.015 & 0.086 & 0.029 \\
& & \checkmark & -- & Cross-Attn. & -- & \textbf{0.025} & \textbf{0.018} & 0.042 & \textbf{0.023} & 0.089 & \textbf{0.035} \\
& & \checkmark & -- & FiLM & -- & 0.022 & 0.012 & \textbf{0.049} & 0.021 & \textbf{0.096} & 0.032 \\
\cmidrule(r){3-12}
& & \checkmark & Concat & Concat & Prompt & 0.034 & \textbf{0.025} & 0.060 & \textbf{0.033} & 0.094 & 0.042 \\
& & \checkmark & Concat & Concat & Embed & 0.022 & 0.014 & 0.045 & 0.022 & 0.091 & 0.033 \\
& & \checkmark & Concat & Concat & -- & \textbf{0.038} & 0.023 & \textbf{0.061} & 0.031 & \textbf{0.114} & \textbf{0.044} \\
& & \checkmark & Weighted Avg. & Weighted Avg. & -- & 0.028 & 0.015 & 0.060 & 0.025 & 0.111 & 0.038 \\
& & \checkmark & Cross-Attn. & Cross-Attn. & -- & 0.022 & 0.013 & 0.042 & 0.020 & 0.094 & 0.033 \\
& & \checkmark & FiLM & FiLM & -- & 0.011 & 0.006 & 0.045 & 0.017 & 0.089 & 0.027 \\
\cmidrule(lr){2-12}
& \multirow{17}{*}{BERT4Rec}
& \checkmark & -- & -- & -- & 0.020 & 0.014 & 0.042 & 0.021 & 0.089 & 0.033 \\
& & \checkmark & -- & -- & Embed & 0.017 & 0.010 & 0.045 & 0.019 & 0.086 & 0.029 \\
& & \checkmark & -- & -- & Prompt & \textbf{0.038} & \textbf{0.024} & \textbf{0.056} & \textbf{0.030} & \textbf{0.092} & \textbf{0.039} \\
\cmidrule(r){3-12}
& & \checkmark & Concat & -- & -- & 0.027 & 0.014 & 0.042 & 0.019 & 0.096 & 0.033 \\
& & \checkmark & Weighted Avg. & -- & -- & 0.020 & 0.012 & 0.045 & 0.020 & 0.096 & 0.032 \\
& & \checkmark & Cross-Attn. & -- & -- & \textbf{0.028} & \textbf{0.016} & \textbf{0.053} & \textbf{0.024} & 0.091 & 0.033 \\
& & \checkmark & FiLM & -- & -- & 0.022 & 0.013 & 0.049 & 0.021 & \textbf{0.099} & \textbf{0.034} \\
\cmidrule(r){3-12}
& & \checkmark & -- & Concat & -- & 0.014 & 0.007 & 0.036 & 0.014 & 0.082 & 0.026 \\
& & \checkmark & -- & Weighted Avg. & -- & \textbf{0.024} & 0.016 & 0.041 & 0.022 & 0.094 & 0.035 \\
& & \checkmark & -- & Cross-Attn. & -- & \textbf{0.024} & \textbf{0.017} & 0.047 & \textbf{0.025} & \textbf{0.108} & \textbf{0.040} \\
& & \checkmark & -- & FiLM & -- & \textbf{0.024} & 0.015 & \textbf{0.053} & 0.024 & \textbf{0.108} & 0.038 \\
\cmidrule(r){3-12}
& & \checkmark & Concat & Concat & Prompt & 0.019 & 0.013 & 0.044 & 0.021 & 0.092 & 0.034 \\
& & \checkmark & Concat & Concat & Embed & 0.024 & 0.015 & 0.041 & 0.020 & 0.099 & 0.035 \\
& & \checkmark & Concat & Concat & -- & 0.033 & \textbf{0.022} & \textbf{0.075} & \textbf{0.036} & \textbf{0.124} & \textbf{0.048} \\
& & \checkmark & Weighted Avg. & Weighted Avg. & -- & 0.025 & 0.016 & 0.053 & 0.025 & 0.105 & 0.037 \\
& & \checkmark & Cross-Attn. & Cross-Attn. & -- & \textbf{0.034} & \textbf{0.022} & 0.056 & 0.029 & 0.102 & 0.040 \\
& & \checkmark & FiLM & FiLM & -- & 0.024 & 0.013 & 0.049 & 0.021 & 0.078 & 0.029 \\
\cmidrule(lr){2-12}
& \multirow{17}{*}{GRU4Rec}
& \checkmark & -- & -- & -- & 0.025 & \textbf{0.015} & 0.044 & 0.021 & 0.088 & 0.032 \\
& & \checkmark & -- & -- & Embed & \textbf{0.027} & \textbf{0.015} & \textbf{0.064} & \textbf{0.027} & 0.118 & 0.040 \\
& & \checkmark & -- & -- & Prompt & 0.024 & 0.013 & 0.055 & 0.023 & \textbf{0.127} & \textbf{0.041} \\
\cmidrule(r){3-12}
& & \checkmark & Concat & -- & -- & 0.025 & 0.014 & \textbf{0.050} & \textbf{0.022} & \textbf{0.100} & \textbf{0.035} \\
& & \checkmark & Weighted Avg. & -- & -- & 0.024 & \textbf{0.015} & 0.042 & 0.021 & 0.080 & 0.030 \\
& & \checkmark & Cross-Attn. & -- & -- & 0.025 & \textbf{0.015} & 0.042 & 0.021 & 0.097 & 0.034 \\
& & \checkmark & FiLM & -- & -- & \textbf{0.028} & 0.014 & 0.045 & 0.019 & 0.096 & 0.032 \\
\cmidrule(r){3-12}
& & \checkmark & -- & Concat & -- & \textbf{0.034} & \textbf{0.019} & \textbf{0.056} & \textbf{0.026} & \textbf{0.100} & 0.037 \\
& & \checkmark & -- & Weighted Avg. & -- & 0.027 & 0.014 & 0.049 & 0.022 & \textbf{0.119} & \textbf{0.039} \\
& & \checkmark & -- & Cross-Attn. & -- & 0.017 & 0.010 & 0.033 & 0.015 & 0.086 & 0.028 \\
& & \checkmark & -- & FiLM & -- & 0.008 & 0.005 & 0.036 & 0.014 & 0.097 & 0.029 \\
\cmidrule(r){3-12}
& & \checkmark & Concat & Concat & Prompt & \textbf{0.025} & \textbf{0.018} & 0.052 & \textbf{0.027} & 0.096 & 0.038 \\
& & \checkmark & Concat & Concat & Embed & 0.019 & 0.013 & 0.038 & 0.019 & 0.099 & 0.034 \\
& & \checkmark & Concat & Concat & -- & \textbf{0.025} & 0.016 & 0.045 & 0.022 & 0.092 & 0.034 \\
& & \checkmark & Weighted Avg. & Weighted Avg. & -- & 0.019 & 0.012 & 0.045 & 0.020 & 0.100 & 0.034 \\
& & \checkmark & Cross-Attn. & Cross-Attn. & -- & \textbf{0.025} & 0.017 & \textbf{0.053} & 0.026 & \textbf{0.116} & \textbf{0.041} \\
& & \checkmark & FiLM & FiLM & -- & \textbf{0.025} & 0.014 & 0.052 & 0.022 & 0.108 & 0.036 \\
\bottomrule
\end{tabular}
}
\end{table*}

\begin{table*}[t]
\centering
\scriptsize
\setlength{\tabcolsep}{3pt}
\caption{Zero-shot results for LLaMA-2-13B.}
\label{tab:LLaMa-2-13B-zeroshot-results}
\resizebox{0.85\textwidth}{!}{
\begin{tabular}{cllcccccccccc}
\toprule
\multirow{2}{*}{LLM}
& \multirow{2}{*}{Seq. Model}
& \multirow{2}{*}{ID}
& \multirow{2}{*}{Audio}
& \multirow{2}{*}{Lyric}
& \multirow{2}{*}{CR}
& \multicolumn{2}{c}{@5}
& \multicolumn{2}{c}{@10}
& \multicolumn{2}{c}{@20} \\
\cmidrule(lr){7-8} \cmidrule(lr){9-10} \cmidrule(lr){11-12}
& & & & & & Recall & NDCG & Recall & NDCG & Recall & NDCG \\
\midrule

\multirow{51}{*}{\rotatebox[origin=c]{90}{LLaMA-2-13B}}
& \multirow{17}{*}{SASRec}
& \checkmark & -- & -- & -- & \textbf{0.030} & \textbf{0.016} & \textbf{0.061} & \textbf{0.026} & \textbf{0.127} & \textbf{0.042} \\
& & \checkmark & -- & -- & Embed & 0.022 & 0.012 & 0.053 & 0.022 & 0.094 & 0.032 \\
& & \checkmark & -- & -- & Prompt & 0.025 & 0.015 & 0.049 & 0.022 & 0.099 & 0.035 \\
& & \checkmark & Concat & -- & -- & 0.025 & \textbf{0.016} & 0.056 & \textbf{0.026} & 0.107 & 0.039 \\
& & \checkmark & Weighted Avg. & -- & -- & 0.019 & 0.009 & 0.049 & 0.018 & 0.096 & 0.030 \\
& & \checkmark & Cross-Attn. & -- & -- & 0.025 & 0.015 & 0.058 & 0.025 & 0.096 & 0.035 \\
& & \checkmark & FiLM & -- & -- & 0.025 & \textbf{0.016} & 0.044 & 0.021 & 0.086 & 0.032 \\
\cmidrule(r){3-12}
& & \checkmark & -- & Concat & -- & \textbf{0.022} & \textbf{0.015} & 0.041 & \textbf{0.021} & 0.100 & \textbf{0.036} \\
& & \checkmark & -- & Weighted Avg. & -- & 0.017 & 0.010 & 0.036 & 0.016 & 0.080 & 0.027 \\
& & \checkmark & -- & Cross-Attn. & -- & 0.020 & 0.012 & \textbf{0.047} & 0.020 & \textbf{0.102} & 0.034 \\
& & \checkmark & -- & FiLM & -- & 0.016 & 0.008 & 0.034 & 0.015 & 0.075 & 0.025 \\
\cmidrule(r){3-12}
& & \checkmark & Concat & Concat & Prompt & \textbf{0.030} & 0.016 & \textbf{0.058} & 0.025 & 0.096 & 0.035 \\
& & \checkmark & Concat & Concat & Embed & 0.028 & \textbf{0.019} & 0.055 & \textbf{0.027} & \textbf{0.107} & \textbf{0.040} \\
& & \checkmark & Concat & Concat & -- & 0.024 & 0.015 & 0.050 & 0.023 & 0.102 & 0.037 \\
& & \checkmark & Weighted Avg. & Weighted Avg. & -- & 0.028 & 0.015 & 0.050 & 0.022 & 0.103 & 0.036 \\
& & \checkmark & Cross-Attn. & Cross-Attn. & -- & 0.019 & 0.010 & 0.045 & 0.019 & 0.075 & 0.026 \\
& & \checkmark & FiLM & FiLM & -- & 0.024 & 0.013 & 0.049 & 0.021 & 0.091 & 0.032 \\
\cmidrule(lr){2-12}
& \multirow{17}{*}{BERT4Rec}
& \checkmark & -- & -- & -- & 0.024 & 0.014 & 0.056 & 0.025 & 0.108 & 0.038 \\
& & \checkmark & -- & -- & Embed & 0.027 & 0.016 & 0.053 & 0.025 & 0.110 & 0.039 \\
& & \checkmark & -- & -- & Prompt & 0.031 & 0.017 & \textbf{0.064} & 0.027 & 0.129 & \textbf{0.044} \\
& & \checkmark & Concat & -- & -- & 0.033 & 0.017 & 0.053 & 0.024 & 0.108 & 0.037 \\
& & \checkmark & Weighted Avg. & -- & -- & 0.009 & 0.005 & 0.034 & 0.013 & 0.075 & 0.024 \\
& & \checkmark & Cross-Attn. & -- & -- & 0.033 & \textbf{0.020} & 0.060 & 0.028 & 0.102 & 0.039 \\
& & \checkmark & FiLM & -- & -- & \textbf{0.034} & \textbf{0.020} & 0.061 & \textbf{0.028} & \textbf{0.122} & \textbf{0.044} \\
\cmidrule(r){3-12}
& & \checkmark & -- & Concat & -- & \textbf{0.036} & \textbf{0.022} & \textbf{0.060} & \textbf{0.029} & \textbf{0.118} & \textbf{0.113} \\
& & \checkmark & -- & Weighted Avg. & -- & 0.028 & 0.014 & 0.050 & 0.021 & 0.103 & 0.035 \\
& & \checkmark & -- & Cross-Attn. & -- & 0.017 & 0.011 & 0.047 & 0.020 & 0.082 & 0.029 \\
& & \checkmark & -- & FiLM & -- & 0.016 & 0.007 & 0.036 & 0.014 & 0.096 & 0.029 \\
\cmidrule(r){3-12}
& & \checkmark & Concat & Concat & Prompt & 0.020 & 0.012 & 0.056 & 0.024 & 0.097 & 0.034 \\
& & \checkmark & Concat & Concat & Embed & 0.031 & 0.018 & 0.053 & 0.025 & 0.097 & 0.036 \\
& & \checkmark & Concat & Concat & -- & \textbf{0.038} & \textbf{0.021} & \textbf{0.063} & \textbf{0.029} & \textbf{0.122} & \textbf{0.044} \\
& & \checkmark & Weighted Avg. & Weighted Avg. & -- & 0.024 & 0.016 & 0.042 & 0.022 & 0.088 & 0.033 \\
& & \checkmark & Cross-Attn. & Cross-Attn. & -- & 0.027 & 0.016 & \textbf{0.063} & 0.027 & 0.108 & 0.039 \\
& & \checkmark & FiLM & FiLM & -- & 0.020 & 0.011 & 0.056 & 0.022 & 0.108 & 0.035 \\
\cmidrule(lr){2-12}
& \multirow{17}{*}{GRU4Rec}
& \checkmark & -- & -- & -- & 0.030 & 0.019 & 0.061 & 0.029 & \textbf{0.125} & 0.045 \\
& & \checkmark & -- & -- & Embed & \textbf{0.034} & \textbf{0.023} & 0.060 & 0.031 & \textbf{0.125} & \textbf{0.048} \\
& & \checkmark & -- & -- & Prompt & 0.038 & 0.025 & 0.058 & 0.032 & 0.113 & 0.045 \\
& & \checkmark & Concat & -- & -- & 0.027 & 0.013 & 0.045 & 0.019 & 0.102 & 0.033 \\
& & \checkmark & Weighted Avg. & -- & -- & 0.019 & 0.010 & 0.047 & 0.020 & 0.108 & 0.035 \\
& & \checkmark & Cross-Attn. & -- & -- & 0.020 & 0.012 & 0.042 & 0.019 & 0.092 & 0.031 \\
& & \checkmark & FiLM & -- & -- & 0.031 & 0.022 & \textbf{0.064} & \textbf{0.032} & 0.121 & 0.047 \\
\cmidrule(r){3-12}
& & \checkmark & -- & Concat & -- & 0.028 & 0.016 & 0.052 & 0.023 & 0.089 & 0.032 \\
& & \checkmark & -- & Weighted Avg. & -- & 0.024 & 0.014 & 0.042 & 0.020 & 0.089 & 0.032 \\
& & \checkmark & -- & Cross-Attn. & -- & 0.024 & 0.014 & 0.060 & 0.025 & 0.094 & 0.033 \\
& & \checkmark & -- & FiLM & -- & \textbf{0.036} & \textbf{0.021} & \textbf{0.061} & \textbf{0.029} & \textbf{0.110} & \textbf{0.041} \\
\cmidrule(r){3-12}
& & \checkmark & Concat & Concat & Prompt & \textbf{0.024} & \textbf{0.015} & 0.044 & 0.021 & 0.097 & 0.035 \\
& & \checkmark & Concat & Concat & Embed & 0.020 & 0.012 & 0.049 & 0.021 & 0.113 & 0.037 \\
& & \checkmark & Concat & Concat & -- & 0.016 & 0.010 & 0.049 & 0.020 & 0.108 & 0.035 \\
& & \checkmark & Weighted Avg. & Weighted Avg. & -- & 0.022 & 0.014 & \textbf{0.053} & \textbf{0.024} & \textbf{0.114} & \textbf{0.039} \\
& & \checkmark & Cross-Attn. & Cross-Attn. & -- & 0.014 & 0.007 & 0.049 & 0.018 & 0.100 & 0.031 \\
& & \checkmark & FiLM & FiLM & -- & 0.020 & \textbf{0.015} & 0.041 & 0.021 & 0.080 & 0.031 \\
\bottomrule
\end{tabular}
}
\end{table*}

\begin{table*}[t]
\centering
\scriptsize
\setlength{\tabcolsep}{3pt}
\renewcommand{\arraystretch}{0.92}
\caption{Zero-shot results for Qwen2.5-7B-Instruct.}
\label{tab:qwen-zeroshot-results}
\resizebox{0.9\textwidth}{!}{
\begin{tabular}{c l c c c c cccccc}
\toprule
\multirow{2}{*}{LLM}
& \multirow{2}{*}{Seq. Model}
& \multirow{2}{*}{ID}
& \multirow{2}{*}{Audio}
& \multirow{2}{*}{Lyric}
& \multirow{2}{*}{CR}
& \multicolumn{2}{c}{@5}
& \multicolumn{2}{c}{@10}
& \multicolumn{2}{c}{@20} \\
\cmidrule(lr){7-8} \cmidrule(lr){9-10} \cmidrule(lr){11-12}
& & & & & & Recall & NDCG & Recall & NDCG & Recall & NDCG \\
\midrule

\multirow{39}{*}{\rotatebox{90}{Qwen2.5-7B-Instruct}}

& \multirow{13}{*}{SASRec}
& \checkmark & -- & -- & --
& 0.033 & \textbf{0.020} & 0.056 & 0.027 & 0.100 & 0.038 \\

& & \checkmark & Concat & -- & --
& 0.016 & 0.009 & 0.041 & 0.017 & 0.097 & 0.031 \\

& & \checkmark & Weighted Avg. & -- & --
& 0.020 & 0.013 & 0.049 & 0.022 & 0.100 & 0.034 \\

& & \checkmark & Cross-Attn. & -- & --
& \textbf{0.038} & 0.019 & \textbf{0.069} & \textbf{0.029} & \textbf{0.110} & \textbf{0.040} \\

& & \checkmark & FiLM & -- & --
& 0.024 & 0.013 & 0.044 & 0.019 & 0.094 & 0.032 \\

\cmidrule(r){3-12}
& & \checkmark & -- & Concat & --
& 0.022 & 0.013 & 0.039 & 0.018 & 0.089 & 0.031 \\

& & \checkmark & -- & Weighted Avg. & --
& \textbf{0.034} & \textbf{0.019} & \textbf{0.063} & \textbf{0.028} & 0.094 & 0.035 \\

& & \checkmark & -- & Cross-Attn. & --
& 0.028 & 0.016 & 0.049 & 0.023 & 0.088 & 0.033 \\

& & \checkmark & -- & FiLM & --
& 0.028 & 0.016 & 0.050 & 0.023 & \textbf{0.100} & \textbf{0.036} \\

\cmidrule(r){3-12}
& & \checkmark & Concat & Concat & --
& 0.017 & 0.009 & 0.044 & 0.017 & 0.091 & 0.029 \\

& & \checkmark & Concat & Concat & Prompt
& 0.025 & 0.016 & 0.042 & 0.022 & 0.097 & 0.035 \\

& & \checkmark & Concat & Concat & Embed
& 0.017 & 0.009 & 0.036 & 0.015 & 0.097 & 0.035 \\

& & \checkmark & Weighted Avg. & Weighted Avg. & --
& 0.020 & 0.014 & 0.039 & 0.020 & 0.091 & 0.029 \\

& & \checkmark & Cross-Attn. & Cross-Attn. & --
& \textbf{0.038} & \textbf{0.024} & 0.045 & \textbf{0.026} & \textbf{0.102} & \textbf{0.040} \\

& & \checkmark & FiLM & FiLM & --
& 0.024 & 0.015 & \textbf{0.053} & 0.024 & 0.099 & 0.036 \\

\cmidrule(lr){2-12}

& \multirow{13}{*}{BERT4Rec}

& \checkmark & -- & -- & --
& 0.033 & 0.020 & 0.055 & 0.026 & 0.102 & 0.038 \\

& & \checkmark & Concat & -- & --
& \textbf{0.041} & \textbf{0.025} & \textbf{0.056} & \textbf{0.030} & 0.103 & \textbf{0.042} \\

& & \checkmark & Weighted Avg. & -- & --
& 0.024 & 0.015 & 0.041 & 0.021 & 0.100 & 0.036 \\

& & \checkmark & Cross-Attn. & -- & --
& 0.033 & 0.019 & \textbf{0.056} & 0.026 & 0.103 & 0.038 \\

& & \checkmark & FiLM & -- & --
& 0.024 & 0.014 & 0.047 & 0.021 & \textbf{0.114} & 0.038 \\

\cmidrule(r){3-12}
& & \checkmark & -- & Concat & --
& 0.024 & \textbf{0.017} & 0.050 & \textbf{0.026} & \textbf{0.110} & \textbf{0.041} \\

& & \checkmark & -- & Weighted Avg. & --
& 0.027 & 0.015 & 0.047 & 0.022 & 0.102 & 0.035 \\

& & \checkmark & -- & Cross-Attn. & --
& \textbf{0.028} & 0.015 & \textbf{0.053} & 0.023 & 0.089 & 0.032 \\

& & \checkmark & -- & FiLM & --
& 0.024 & 0.013 & 0.050 & 0.022 & 0.092 & 0.032 \\

\cmidrule(r){3-12}
& & \checkmark & Concat & Concat & --
& \textbf{0.025} & \textbf{0.013} & \textbf{0.052} & \textbf{0.021} & \textbf{0.103} & \textbf{0.034} \\

& & \checkmark & Weighted Avg. & Weighted Avg. & --
& 0.022 & 0.011 & \textbf{0.052} & \textbf{0.021} & 0.097 & 0.032 \\

& & \checkmark & Cross-Attn. & Cross-Attn. & --
& 0.014 & 0.008 & 0.041 & 0.017 & 0.088 & 0.029 \\

& & \checkmark & FiLM & FiLM & --
& 0.014 & 0.010 & 0.047 & 0.020 & 0.100 & \textbf{0.034} \\

\cmidrule(lr){2-12}

& \multirow{13}{*}{GRU4Rec}

& \checkmark & -- & -- & --
& 0.025 & \textbf{0.015} & \textbf{0.058} & \textbf{0.026} & 0.108 & \textbf{0.038} \\

& & \checkmark & Concat & -- & --
& 0.025 & 0.013 & 0.044 & 0.018 & 0.100 & 0.033 \\

& & \checkmark & Weighted Avg. & -- & --
& 0.025 & 0.013 & 0.049 & 0.021 & \textbf{0.113} & 0.037 \\

& & \checkmark & Cross-Attn. & -- & --
& \textbf{0.027} & \textbf{0.015} & 0.050 & 0.022 & 0.103 & 0.035 \\

& & \checkmark & FiLM & -- & --
& 0.025 & 0.014 & 0.045 & 0.020 & 0.094 & 0.032 \\

\cmidrule(r){3-12}
& & \checkmark & -- & Concat & --
& \textbf{0.028} & \textbf{0.018} & 0.058 & \textbf{0.028} & 0.100 & \textbf{0.038} \\

& & \checkmark & -- & Weighted Avg. & --
& 0.022 & 0.013 & 0.053 & 0.023 & \textbf{0.103} & 0.035 \\

& & \checkmark & -- & Cross-Attn. & --
& \textbf{0.028} & 0.016 & \textbf{0.061} & 0.027 & 0.099 & 0.036 \\

& & \checkmark & -- & FiLM & --
& 0.025 & 0.014 & 0.045 & 0.021 & 0.096 & 0.033 \\

\cmidrule(r){3-12}
& & \checkmark & Concat & Concat & --
& \textbf{0.036} & \textbf{0.021} & \textbf{0.066} & \textbf{0.031} & \textbf{0.114} & \textbf{0.043} \\

& & \checkmark & Weighted Avg. & Weighted Avg. & --
& 0.024 & 0.013 & 0.047 & 0.021 & 0.092 & 0.032 \\

& & \checkmark & Cross-Attn. & Cross-Attn. & --
& 0.017 & 0.008 & 0.041 & 0.016 & 0.091 & 0.028 \\

& & \checkmark & FiLM & FiLM & --
& 0.025 & 0.014 & 0.047 & 0.021 & 0.107 & 0.035 \\

\bottomrule
\end{tabular}
}
\end{table*}

\begin{table*}[t]
\centering
\scriptsize
\setlength{\tabcolsep}{3pt}
\caption{Fine-tuned results for LLaMA-2-13B.}
\label{tab:LLaMa-2-13B-finetuned-results}
\resizebox{0.625\textwidth}{!}{
\begin{tabular}{cllcccccccccc}
\toprule
\multirow{2}{*}{LLM}
& \multirow{2}{*}{Seq. Model}
& \multirow{2}{*}{ID}
& \multirow{2}{*}{Audio}
& \multirow{2}{*}{Lyric}
& \multirow{2}{*}{CR}
& \multicolumn{2}{c}{@5}
& \multicolumn{2}{c}{@10}
& \multicolumn{2}{c}{@20} \\
\cmidrule(lr){7-8} \cmidrule(lr){9-10} \cmidrule(lr){11-12}
& & & & & & Recall & NDCG & Recall & NDCG & Recall & NDCG \\
\midrule

\multirow{63}{*}{\rotatebox[origin=c]{90}{LLaMA-2-13B-Fine-tuned}}
& \multirow{21}{*}{SASRec}

& \checkmark & -- & -- & -- & 0.229 & 0.125 & 0.323 & 0.155 & 0.398 & 0.174 \\

& & \checkmark & Concat & -- & -- & 0.223 & 0.122 & 0.323 & 0.154 & 0.401 & 0.174 \\
& & \checkmark & Weighted Avg. & -- & -- & 0.232 & 0.124 & 0.317 & 0.151 & 0.397 & 0.172 \\
& & \checkmark & Weighted Avg. & -- & Prompt & 0.219 & 0.119 & 0.321 & 0.152 & 0.397 & 0.171 \\
& & \checkmark & Cross-Attn. & -- & -- & \textbf{0.240} & \textbf{0.130} & 0.326 & 0.158 & 0.395 & 0.175 \\
& & \checkmark & Cross-Attn. & -- & Prompt & 0.227 & 0.123 & 0.324 & 0.154 & \textbf{0.406} & 0.175 \\
& & \checkmark & FiLM & -- & -- & 0.232 & 0.128 & \textbf{0.334} & \textbf{0.161} & 0.398 & \textbf{0.177} \\
& & \checkmark & FiLM & -- & Prompt & 0.234 & 0.128 & 0.328 & 0.158 & 0.398 & \textbf{0.177} \\
\cmidrule(r){3-12}

& & \checkmark & -- & Concat & -- & 0.229 & 0.127 & 0.321 & \textbf{0.157} & 0.395 & \textbf{0.176} \\
& & \checkmark & -- & Weighted Avg. & -- & 0.208 & 0.114 & 0.317 & 0.149 & 0.390 & 0.168 \\
& & \checkmark & -- & Weighted Avg. & Prompt & \textbf{0.241} & \textbf{0.129} & \textbf{0.326} & 0.156 & 0.398 & 0.175 \\
& & \checkmark & -- & Cross-Attn. & -- & 0.227 & 0.123 & 0.321 & 0.154 & \textbf{0.400} & 0.174 \\
& & \checkmark & -- & FiLM & -- & 0.215 & 0.118 & 0.298 & 0.145 & 0.387 & 0.168 \\
& & \checkmark & -- & FiLM & Prompt & 0.227 & 0.124 & 0.313 & 0.152 & 0.395 & 0.173 \\
\cmidrule(r){3-12}

& & \checkmark & Concat & Concat & -- & 0.234 & 0.127 & 0.317 & 0.154 & \textbf{0.404} & \textbf{0.177} \\
& & \checkmark & Weighted Avg. & Weighted Avg. & -- & 0.230 & 0.125 & 0.313 & 0.152 & 0.395 & 0.173 \\
& & \checkmark & Weighted Avg. & Weighted Avg. & Prompt & 0.227 & 0.125 & 0.321 & 0.155 & 0.393 & 0.173 \\
& & \checkmark & Cross-Attn. & Cross-Attn. & -- & 0.188 & 0.105 & 0.287 & 0.137 & 0.346 & 0.151 \\
& & \checkmark & Cross-Attn. & Cross-Attn. & Prompt & 0.226 & 0.124 & 0.323 & 0.156 & 0.398 & 0.175 \\
& & \checkmark & FiLM & FiLM & -- & \textbf{0.241} & \textbf{0.131} & \textbf{0.326} & \textbf{0.158} & 0.395 & 0.176 \\
& & \checkmark & FiLM & FiLM & Prompt & 0.234 & 0.126 & 0.323 & 0.155 & 0.395 & 0.173 \\
\cmidrule(lr){2-12}

& \multirow{24}{*}{BERT4Rec}

& \checkmark & -- & -- & -- & 0.224 & 0.122 & 0.321 & 0.154 & 0.392 & 0.172 \\
& & \checkmark & -- & -- & Prompt & 0.224 & 0.121 & 0.328 & 0.155 & 0.400 & 0.173 \\

& & \checkmark & Concat & -- & -- & 0.240 & \textbf{0.130} & 0.323 & 0.156 & \textbf{0.404} & \textbf{0.177} \\
& & \checkmark & Concat & -- & Prompt & 0.229 & 0.125 & \textbf{0.329} & \textbf{0.157} & 0.401 & 0.176 \\
& & \checkmark & Weighted Avg. & -- & -- & 0.232 & 0.125 & 0.323 & 0.155 & 0.392 & 0.173 \\
& & \checkmark & Weighted Avg. & -- & Prompt & 0.223 & 0.121 & 0.321 & 0.153 & 0.401 & 0.173 \\
& & \checkmark & Cross-Attn. & -- & -- & 0.234 & 0.125 & 0.315 & 0.152 & 0.395 & 0.172 \\
& & \checkmark & Cross-Attn. & -- & Prompt & 0.229 & 0.126 & 0.323 & 0.156 & 0.403 & \textbf{0.177} \\
& & \checkmark & FiLM & -- & -- & 0.234 & 0.128 & 0.335 & \textbf{0.161} & 0.398 & \textbf{0.177} \\
& & \checkmark & FiLM & -- & Prompt & \textbf{0.235} & 0.128 & 0.329 & 0.158 & 0.398 & 0.176 \\
\cmidrule(r){3-12}

& & \checkmark & -- & Concat & -- & 0.224 & 0.121 & 0.321 & 0.153 & \textbf{0.398} & 0.172 \\
& & \checkmark & -- & Concat & Prompt & 0.224 & 0.121 & 0.313 & 0.149 & \textbf{0.398} & 0.171 \\
& & \checkmark & -- & Weighted Avg. & -- & 0.235 & \textbf{0.127} & 0.318 & \textbf{0.153} & 0.395 & \textbf{0.173} \\
& & \checkmark & -- & Weighted Avg. & Prompt & 0.223 & 0.121 & \textbf{0.323} & \textbf{0.153} & 0.397 & 0.172 \\
& & \checkmark & -- & Cross-Attn. & -- & 0.215 & 0.117 & 0.301 & 0.145 & 0.381 & 0.166 \\
& & \checkmark & -- & Cross-Attn. & Prompt & \textbf{0.237} & 0.126 & 0.310 & 0.150 & 0.395 & 0.172 \\
& & \checkmark & -- & FiLM & -- & 0.226 & 0.123 & 0.318 & \textbf{0.153} & 0.397 & \textbf{0.173} \\
& & \checkmark & -- & FiLM & Prompt & 0.227 & 0.122 & 0.312 & 0.150 & 0.397 & 0.172 \\
\cmidrule(r){3-12}

& & \checkmark & Concat & Concat & -- & 0.232 & 0.127 & 0.318 & 0.155 & 0.393 & 0.174 \\
& & \checkmark & Concat & Concat & Prompt & 0.237 & 0.129 & 0.323 & \textbf{0.156} & 0.400 & 0.176 \\
& & \checkmark & Weighted Avg. & Weighted Avg. & -- & 0.227 & 0.123 & \textbf{0.324} & 0.154 & 0.397 & 0.173 \\
& & \checkmark & Weighted Avg. & Weighted Avg. & Prompt & 0.229 & 0.124 & 0.310 & 0.150 & 0.395 & 0.172 \\
& & \checkmark & Cross-Attn. & Cross-Attn. & -- & 0.230 & 0.124 & \textbf{0.324} & 0.155 & 0.401 & 0.174 \\
& & \checkmark & Cross-Attn. & Cross-Attn. & Prompt & 0.230 & 0.126 & 0.323 & \textbf{0.156} & 0.393 & 0.175 \\
& & \checkmark & FiLM & FiLM & -- & \textbf{0.241} & 0.130 & 0.318 & 0.155 & 0.393 & 0.174 \\
& & \checkmark & FiLM & FiLM & Prompt & 0.238 & \textbf{0.131} & 0.317 & \textbf{0.156} & \textbf{0.406} & \textbf{0.179} \\
\cmidrule(lr){2-12}

& \multirow{26}{*}{GRU4Rec}

& \checkmark & -- & -- & -- & 0.228 & 0.124 & 0.316 & 0.152 & 0.393 & 0.172 \\
& & \checkmark & -- & -- & Prompt & 0.228 & 0.123 & 0.321 & 0.153 & 0.399 & 0.173 \\

& & \checkmark & Concat & -- & -- & 0.227 & 0.124 & \textbf{0.330} & \textbf{0.157} & 0.398 & 0.174 \\
& & \checkmark & Concat & -- & Prompt & 0.231 & 0.124 & 0.318 & 0.152 & 0.399 & 0.172 \\
& & \checkmark & Weighted Avg. & -- & -- & 0.230 & 0.124 & 0.321 & 0.153 & 0.390 & 0.171 \\
& & \checkmark & Weighted Avg. & -- & Prompt & 0.235 & 0.125 & 0.319 & 0.153 & 0.396 & 0.173 \\
& & \checkmark & Cross-Attn. & -- & -- & \textbf{0.239} & \textbf{0.128} & 0.313 & 0.152 & \textbf{0.402} & \textbf{0.175} \\
& & \checkmark & Cross-Attn. & -- & Prompt & 0.227 & 0.125 & 0.319 & 0.155 & 0.398 & \textbf{0.175} \\
& & \checkmark & FiLM & -- & -- & 0.230 & 0.124 & 0.318 & 0.152 & 0.388 & 0.171 \\
& & \checkmark & FiLM & -- & Prompt & 0.230 & 0.126 & 0.319 & 0.155 & 0.398 & \textbf{0.175} \\
\cmidrule(r){3-12}

& & \checkmark & -- & Concat & -- & 0.211 & 0.115 & 0.313 & 0.148 & 0.380 & 0.165 \\
& & \checkmark & -- & Concat & Prompt & 0.206 & 0.112 & 0.294 & 0.140 & 0.379 & 0.161 \\
& & \checkmark & -- & Weighted Avg. & -- & 0.219 & 0.118 & 0.296 & 0.142 & 0.380 & 0.164 \\
& & \checkmark & -- & Weighted Avg. & Prompt & \textbf{0.239} & \textbf{0.128} & \textbf{0.322} & \textbf{0.155} & 0.393 & 0.173 \\
& & \checkmark & -- & Cross-Attn. & -- & 0.217 & 0.118 & 0.307 & 0.148 & 0.391 & 0.170 \\
& & \checkmark & -- & Cross-Attn. & Prompt & 0.230 & 0.124 & \textbf{0.322} & 0.154 & 0.401 & 0.174 \\
& & \checkmark & -- & FiLM & -- & 0.216 & 0.115 & 0.297 & 0.142 & 0.387 & 0.165 \\
& & \checkmark & -- & FiLM & Prompt & 0.222 & 0.123 & 0.311 & 0.153 & \textbf{0.404} & \textbf{0.177} \\
\cmidrule(r){3-12}

& & \checkmark & Concat & Concat & -- & \textbf{0.236} & \textbf{0.129} & 0.327 & 0.158 & \textbf{0.396} & \textbf{0.176} \\
& & \checkmark & Concat & Concat & Prompt & 0.231 & 0.126 & 0.322 & 0.156 & 0.390 & 0.173 \\
& & \checkmark & Weighted Avg. & Weighted Avg. & -- & 0.191 & 0.106 & 0.280 & 0.134 & 0.335 & 0.149 \\
& & \checkmark & Weighted Avg. & Weighted Avg. & Prompt & 0.224 & 0.122 & 0.321 & 0.153 & 0.393 & 0.172 \\
& & \checkmark & Cross-Attn. & Cross-Attn. & -- & 0.231 & 0.124 & 0.318 & 0.152 & 0.391 & 0.171 \\
& & \checkmark & Cross-Attn. & Cross-Attn. & Prompt & 0.227 & 0.122 & 0.326 & 0.154 & 0.399 & 0.173 \\
& & \checkmark & FiLM & FiLM & -- & 0.238 & \textbf{0.129} & \textbf{0.330} & \textbf{0.159} & 0.388 & 0.174 \\
& & \checkmark & FiLM & FiLM & Prompt & \textbf{0.236} & 0.127 & 0.326 & 0.155 & \textbf{0.396} & 0.173 \\
\bottomrule
\end{tabular}
}
\end{table*}

\begin{table*}[t]
\centering
\scriptsize
\setlength{\tabcolsep}{3pt}
\caption{Fine-tuned results for Qwen2.5-7B-Instruct.}
\label{tab:qwen-finetuned-results}
\resizebox{0.655\textwidth}{!}{
\begin{tabular}{cllcccccccccc}
\toprule
\multirow{2}{*}{LLM}
& \multirow{2}{*}{Seq. Model}
& \multirow{2}{*}{ID}
& \multirow{2}{*}{Audio}
& \multirow{2}{*}{Lyric}
& \multirow{2}{*}{CR}
& \multicolumn{2}{c}{@5}
& \multicolumn{2}{c}{@10}
& \multicolumn{2}{c}{@20} \\
\cmidrule(lr){7-8} \cmidrule(lr){9-10} \cmidrule(lr){11-12}
& & & & & & Recall & NDCG & Recall & NDCG & Recall & NDCG \\
\midrule

\multirow{67}{*}{\rotatebox[origin=c]{90}{Qwen2.5-7B-Instruct-Fine-tuned}}
& \multirow{17}{*}{SASRec}

& \checkmark & -- & -- & -- & 0.188 & 0.102 & 0.268 & 0.128 & 0.346 & 0.148 \\

& & \checkmark & Concat & -- & -- & \textbf{0.226} & \textbf{0.123} & \textbf{0.326} & \textbf{0.155} & 0.398 & \textbf{0.173} \\
& & \checkmark & Concat & -- & Prompt & 0.224 & 0.120 & 0.312 & 0.149 & 0.381 & 0.167 \\
& & \checkmark & Weighted Avg. & -- & -- & 0.210 & 0.115 & 0.313 & 0.149 & 0.386 & 0.167 \\
& & \checkmark & Cross-Attn. & -- & -- & 0.224 & 0.121 & 0.323 & 0.153 & \textbf{0.400} & \textbf{0.173} \\
& & \checkmark & FiLM & -- & -- & 0.221 & 0.118 & 0.317 & 0.150 & 0.392 & 0.169 \\
\cmidrule(r){3-12}

& & \checkmark & -- & Concat & -- & \textbf{0.227} & \textbf{0.123} & \textbf{0.324} & \textbf{0.155} & 0.390 & \textbf{0.172} \\
& & \checkmark & -- & Concat & Prompt & 0.215 & 0.117 & 0.312 & 0.148 & 0.386 & 0.167 \\
& & \checkmark & -- & Weighted Avg. & -- & 0.212 & 0.115 & 0.288 & 0.140 & 0.389 & 0.166 \\
& & \checkmark & -- & Weighted Avg. & Prompt & 0.213 & 0.115 & 0.312 & 0.147 & 0.390 & 0.167 \\
& & \checkmark & -- & Cross-Attn. & -- & 0.216 & 0.116 & 0.293 & 0.140 & 0.381 & 0.162 \\
& & \checkmark & -- & FiLM & -- & 0.218 & 0.118 & 0.304 & 0.146 & \textbf{0.403} & 0.171 \\
\cmidrule(r){3-12}

& & \checkmark & Concat & Concat & -- & \textbf{0.229} & \textbf{0.124} & 0.313 & \textbf{0.152} & 0.398 & \textbf{0.173} \\
& & \checkmark & Concat & Concat & Prompt & 0.199 & 0.104 & 0.273 & 0.128 & 0.340 & 0.145 \\
& & \checkmark & Weighted Avg. & Weighted Avg. & -- & 0.224 & 0.120 & 0.306 & 0.146 & \textbf{0.400} & 0.170 \\
& & \checkmark & Cross-Attn. & Cross-Attn. & -- & 0.224 & 0.122 & \textbf{0.318} & \textbf{0.152} & 0.398 & 0.172 \\
& & \checkmark & FiLM & FiLM & -- & 0.216 & 0.121 & 0.317 & 0.153 & 0.384 & 0.170 \\

\cmidrule(lr){2-12}
& \multirow{25}{*}{BERT4Rec}

& \checkmark & -- & -- & -- & 0.224 & 0.122 & 0.310 & 0.150 & 0.393 & 0.171 \\
& & \checkmark & -- & -- & Prompt & 0.219 & 0.118 & 0.307 & 0.150 & \textbf{0.401} & 0.170 \\

& & \checkmark & Concat & -- & -- & \textbf{0.235} & \textbf{0.127} & 0.310 & \textbf{0.151} & 0.390 & \textbf{0.171} \\
& & \checkmark & Concat & -- & Prompt & 0.218 & 0.119 & 0.309 & 0.148 & 0.390 & 0.170 \\
& & \checkmark & Weighted Avg. & -- & -- & 0.212 & 0.116 & 0.307 & 0.146 & 0.392 & 0.168 \\
& & \checkmark & Weighted Avg. & -- & Prompt & 0.223 & 0.118 & 0.313 & 0.148 & 0.389 & 0.167 \\
& & \checkmark & Cross-Attn. & -- & -- & 0.219 & 0.118 & 0.320 & \textbf{0.151} & 0.386 & 0.168 \\
& & \checkmark & Cross-Attn. & -- & Prompt & 0.216 & 0.118 & \textbf{0.323} & 0.152 & \textbf{0.395} & \textbf{0.171} \\
& & \checkmark & FiLM & -- & -- & 0.208 & 0.113 & 0.304 & 0.144 & 0.389 & 0.166 \\
& & \checkmark & FiLM & -- & Prompt & \textbf{0.238} & \textbf{0.124} & 0.320 & \textbf{0.151} & \textbf{0.395} & 0.170 \\
\cmidrule(r){3-12}

& & \checkmark & -- & Concat & -- & 0.224 & 0.120 & 0.309 & 0.148 & 0.390 & 0.169 \\
& & \checkmark & -- & Concat & Prompt & 0.219 & 0.118 & 0.313 & 0.149 & 0.393 & 0.169 \\
& & \checkmark & -- & Weighted Avg. & -- & 0.213 & 0.116 & 0.299 & 0.144 & 0.389 & 0.167 \\
& & \checkmark & -- & Weighted Avg. & Prompt & 0.212 & 0.114 & 0.307 & 0.145 & 0.386 & 0.165 \\
& & \checkmark & -- & Cross-Attn. & -- & \textbf{0.226} & \textbf{0.121} & 0.301 & 0.145 & 0.387 & 0.168 \\
& & \checkmark & -- & Cross-Attn. & Prompt & 0.208 & 0.112 & 0.310 & 0.145 & 0.390 & 0.165 \\
& & \checkmark & -- & FiLM & -- & 0.223 & 0.120 & \textbf{0.315} & \textbf{0.150} & \textbf{0.397} & \textbf{0.171} \\
& & \checkmark & -- & FiLM & Prompt & 0.200 & 0.106 & 0.293 & 0.136 & 0.384 & 0.159 \\
\cmidrule(r){3-12}

& & \checkmark & Concat & Concat & -- & 0.219 & 0.118 & 0.313 & 0.149 & 0.390 & 0.168 \\
& & \checkmark & Concat & Concat & Prompt & 0.215 & 0.116 & 0.309 & 0.146 & 0.384 & 0.165 \\
& & \checkmark & Weighted Avg. & Weighted Avg. & -- & 0.221 & 0.121 & 0.310 & 0.150 & \textbf{0.398} & \textbf{0.172} \\
& & \checkmark & Weighted Avg. & Weighted Avg. & Prompt & 0.207 & 0.111 & 0.298 & 0.141 & 0.379 & 0.162 \\
& & \checkmark & Cross-Attn. & Cross-Attn. & -- & \textbf{0.226} & \textbf{0.123} & 0.315 & \textbf{0.152} & 0.393 & \textbf{0.172} \\
& & \checkmark & Cross-Attn. & Cross-Attn. & Prompt & 0.210 & 0.116 & \textbf{0.320} & \textbf{0.152} & 0.386 & 0.168 \\
& & \checkmark & FiLM & FiLM & -- & 0.216 & 0.117 & 0.300 & 0.146 & 0.390 & 0.167 \\
& & \checkmark & FiLM & FiLM & Prompt & 0.207 & 0.108 & 0.290 & 0.135 & 0.380 & 0.157 \\

\cmidrule(lr){2-12}
& \multirow{25}{*}{GRU4Rec}

& \checkmark & -- & -- & -- & 0.219 & 0.118 & 0.313 & 0.148 & 0.401 & 0.170 \\
& & \checkmark & -- & -- & Prompt & 0.224 & 0.120 & 0.326 & 0.153 & 0.394 & 0.171 \\

& & \checkmark & Concat & -- & -- & \textbf{0.230} & \textbf{0.124} & 0.315 & 0.151 & 0.387 & 0.169 \\
& & \checkmark & Concat & -- & Prompt & 0.221 & 0.118 & 0.318 & 0.150 & 0.388 & 0.168 \\
& & \checkmark & Weighted Avg. & -- & -- & \textbf{0.230} & 0.121 & \textbf{0.322} & 0.151 & \textbf{0.396} & 0.170 \\
& & \checkmark & Weighted Avg. & -- & Prompt & 0.211 & 0.115 & 0.316 & 0.149 & 0.391 & 0.168 \\
& & \checkmark & Cross-Attn. & -- & -- & 0.205 & 0.106 & 0.264 & 0.125 & 0.329 & 0.141 \\
& & \checkmark & Cross-Attn. & -- & Prompt & 0.216 & 0.118 & 0.318 & \textbf{0.152} & 0.393 & \textbf{0.171} \\
& & \checkmark & FiLM & -- & -- & 0.197 & 0.108 & 0.275 & 0.133 & 0.344 & 0.151 \\
& & \checkmark & FiLM & -- & Prompt & 0.211 & 0.115 & 0.319 & 0.150 & \textbf{0.396} & 0.170 \\
\cmidrule(r){3-12}

& & \checkmark & -- & Concat & -- & \textbf{0.231} & \textbf{0.122} & 0.311 & 0.148 & \textbf{0.398} & \textbf{0.170} \\
& & \checkmark & -- & Concat & Prompt & 0.213 & 0.115 & 0.307 & 0.145 & 0.385 & 0.165 \\
& & \checkmark & -- & Weighted Avg. & -- & 0.216 & 0.118 & \textbf{0.315} & \textbf{0.150} & 0.396 & \textbf{0.170} \\
& & \checkmark & -- & Weighted Avg. & Prompt & 0.211 & 0.112 & 0.308 & 0.144 & 0.385 & 0.163 \\
& & \checkmark & -- & Cross-Attn. & -- & 0.210 & 0.113 & 0.305 & 0.145 & 0.390 & 0.166 \\
& & \checkmark & -- & Cross-Attn. & Prompt & 0.203 & 0.109 & 0.299 & 0.141 & 0.385 & 0.163 \\
& & \checkmark & -- & FiLM & -- & 0.221 & 0.120 & 0.307 & 0.147 & 0.385 & 0.167 \\
& & \checkmark & -- & FiLM & Prompt & 0.211 & 0.115 & 0.300 & 0.143 & 0.382 & 0.164 \\
\cmidrule(r){3-12}

& & \checkmark & Concat & Concat & -- & 0.224 & 0.120 & \textbf{0.321} & \textbf{0.151} & 0.393 & 0.169 \\
& & \checkmark & Concat & Concat & Prompt & 0.222 & 0.119 & \textbf{0.321} & \textbf{0.151} & 0.387 & 0.168 \\
& & \checkmark & Weighted Avg. & Weighted Avg. & -- & \textbf{0.230} & \textbf{0.122} & 0.313 & 0.149 & 0.390 & 0.168 \\
& & \checkmark & Weighted Avg. & Weighted Avg. & Prompt & 0.217 & 0.116 & 0.291 & 0.140 & 0.387 & 0.164 \\
& & \checkmark & Cross-Attn. & Cross-Attn. & -- & 0.219 & 0.119 & 0.313 & 0.149 & 0.379 & 0.166 \\
& & \checkmark & Cross-Attn. & Cross-Attn. & Prompt & 0.214 & 0.115 & 0.319 & 0.150 & 0.388 & 0.168 \\
& & \checkmark & FiLM & FiLM & -- & 0.219 & 0.119 & 0.307 & 0.147 & \textbf{0.396} & \textbf{0.170} \\
& & \checkmark & FiLM & FiLM & Prompt & 0.197 & 0.107 & 0.275 & 0.132 & 0.366 & 0.155 \\
\bottomrule
\end{tabular}
}
\end{table*}

\subsection{Impact of Metadata Features}
We study the contribution of semantic metadata under zero-shot inference across LLM backbones, ItemID backbones, and six metadata configurations that vary the source of metadata:
Spotify-derived, MGPHot-derived, or their combination, and whether completion-ratio (CR) signals are incorporated.

Table~\ref{tab:zero-shot-all} report Recall@20, NDCG@10, and MRR@10 for all configurations. Per-cell winner counts are roughly balanced across the 18 experiments. Among all the configurations, LLaMA-3-70B wins 4, LLaMA-2-7B wins 9, and Qwen2.5-7B-Instruct wins 5 times. This suggests that, in a frozen zero-shot setting, larger LLMs do not improve ranking quality.

Across ItemID backbones, the best NDCG@10 values are tightly grouped, with 0.024 for SASRec, 0.022 for BERT4Rec, and 0.021 for GRU4Rec, indicating that the backbone matters less once the LLM front-end and metadata are added.

A head-to-head comparison isolates where the useful signal comes from. The ID-only LLaMA-3-70B baseline with SASRec reaches NDCG@10 of 0.023. Injecting all Spotify columns reduces performance to 0.018, a 22\% drop, while injecting all 58 MGPHot perceptual annotations slightly improves it to 0.024. Naively concatenating both sources into a 76-column prompt reduces it further to 0.016, a 30\% drop.

We observe the same pattern across all experiments. MGPHot produces the strongest overall Recall@20 of 0.122 and NDCG@10 values of 0.024, suggesting that LLM-generated perceptual annotations provide useful semantic signal for both retrieval and ranking. Spotify remains competitive, especially with completion ratio, where LLaMA-2-7B with SASRec reaches Recall@20 of 0.111 and NDCG@10 of 0.023. Combining Spotify and MGPHot (S+M) does not help, with best Recall@20 of 0.114 and NDCG@10 of 0.022, both below MGPHot. Richer metadata is not automatically better, since structure and interpretability matter more than the number of input fields. In particular, descriptive perceptual labels such as ``Cut Time Feel'' and ``String Ensemble'' may be easier for LLMs to interpret than a long concatenation of numeric fields.

MRR@10 is less differentiated, with the best value of 0.013 reached by Spotify, MGPHot, and S+M alike, suggesting that MGPHot mainly improves broader retrieval while first-relevant-item ranking remains comparable. LLM-generated metadata thus provides the dominant signal, with a small residual from Spotify, while uncurated injection of either source, especially their union, adds more noise than useful signal.

\subsection{Ablation Study}
We ablate embedding models, fusion strategies, item backbones, and completion-ratio injection under the same train/validation/test splits and evaluation protocol.

\medskip\noindent\textbf{\textit{Zero-Shot Results.}}
Tables~\ref{tab:LLaMa-3-70B-zeroshot-results}--\ref{tab:qwen-zeroshot-results} report zero-shot results for LLaMA-3-70B, LLaMA-2-13B, and Qwen2.5-7B-Instruct. No LLM receives task-specific fine-tuning, and content features are injected only at inference. Multimodal augmentation yields modest but inconsistent gains over ItemID-only baselines, and the benefit depends on the LLM backbone.

LLaMA-3-70B shows a mixed pattern, where multimodal and engagement-aware features help only in selected configurations. With SASRec, Audio Cross-Attention improves Recall@10 from 0.049 to 0.063, and Audio FiLM raises NDCG@20 from 0.036 to 0.045. With BERT4Rec, Audio + Lyric concatenation gives the strongest result, improving Recall@10 from 0.044 to 0.075 and NDCG@20 from 0.034 to 0.048. GRU4Rec benefits more from completion-ratio signals than from content fusion, where CR embedding improves Recall@10 from 0.044 to 0.064 and CR prompt injection improves Recall@20 from 0.088 to 0.127. Despite its larger scale, LLaMA-3-70B is not uniformly stronger than the smaller models, and its best gains depend heavily on the item backbone and fusion strategy.

LLaMA-2-13B gains little from content features, which in several cases hurt performance. With SASRec, the ItemID-only configuration is the strongest, with Recall@10 of 0.061 and Recall@20 of 0.127. With GRU4Rec, audio FiLM roughly matches the ItemID baseline with Recall@10 of 0.064 and NDCG@20 of 0.047, while combined configurations fall below it. BERT4Rec is the exception, where audio-lyric concatenation improves Recall@10 from 0.056 to 0.063, and prompt-injected completion-ratio (CR) signals raise Recall@20 from 0.108 to 0.129. Its stronger zero-shot sequential reasoning likely leaves less headroom for content-based improvement.

Qwen2.5-7B-Instruct is more responsive to content augmentation. With SASRec, audio cross-attention improves Recall@10 from 0.056 to 0.069 and NDCG@10 from 0.027 to 0.029. With GRU4Rec, audio-lyric concatenation gives the strongest zero-shot result, with Recall@10 of 0.066 and NDCG@20 of 0.043, versus 0.058 and 0.038 for the ItemID-only baseline. Gains are not uniform, since concatenation with SASRec falls below the baseline and audio-lyric cross-attention underperforms on all three item backbones. Qwen's zero-shot use of content thus depends on fusion strategy and item encoder.

Overall, the strongest zero-shot configuration for each model uses content modalities or completion-ratio signals, provided they align with the item backbone and fusion mechanism.

\medskip\noindent\textbf{\textit{Fine-Tuned Results.}}
As shown in Tables \ref{tab:LLaMa-2-13B-finetuned-results} and ~\ref{tab:qwen-finetuned-results}, fine-tuning produces substantial gains over zero-shot performance across both
LLM backbones.
LLaMA-2-13B improves the Recall@10 to 0.323, and fine-tuned Qwen2.5-7B-Instruct achieves a Recall@10 of 0.268 under the ItemID-only configuration compared to 0.056 in the zero-shot setting. More importantly, fine-tuning also stabilizes the contribution
of multimodal content features: whereas zero-shot content augmentation produced inconsistent gains and frequently degraded below the ItemID-only baseline, the best content-augmented configuration improves over the fine-tuned ItemID-only baseline for every backbone, although some fusion choices fall below it. This consistency suggests that the audio and lyric features provided by the benchmark are learnable signals
that task-specific adaptation can reliably exploit, even when they provide only limited benefit in frozen zero-shot conditions.

The magnitude of content feature gains remains moderate relative to the fine-tuned ID baseline. For LLaMA-2-13B, gains over the ID-only baseline are smaller in absolute terms given its already stronger sequential baseline, which achieves Recall@10 of 0.323, and audio FiLM gives the best SASRec result, with Recall@10 of 0.334 and NDCG@10 of 0.161, while lyric FiLM drops to 0.298. For Qwen2.5-7B-Instruct, audio concatenation gives the largest single-modality improvement, with Recall@10 rising from 0.268 to 0.326 and NDCG@10 from 0.128 to 0.155, while lyric FiLM leads with Recall@20 of 0.403. Audio-lyric combinations are comparable at K=20 but do not always surpass the best single-modality ones at K=10, suggesting diminishing returns from naive fusion with this backbone.

Across item backbones, content features help most when the ItemID baseline is weak. With fine-tuned Qwen2.5-7B-Instruct, SASRec starts lowest at Recall@10 of 0.268 and gains the most, where audio concatenation raises it to 0.326, while BERT4Rec and GRU4Rec start near 0.310 and audio and lyric features add at most 0.010. With 638 test users, one standard error at Recall@10 is 0.018, and only the SASRec gains clear it. Completion-ratio prompts add up to 0.013 for BERT4Rec and GRU4Rec, also within one standard error. Fusion choice matters more than modality choice. Concatenation and weighted sum are stable across backbones, while cross-attention and FiLM drop GRU4Rec to 0.264 and 0.275, below its 0.313 baseline. This is consistent with fusion parameters overfitting when the encoder is already strong, which also explains why audio plus lyric does not consistently beat audio alone.

\section{Conclusion}

In this work, we construct a multimodal benchmark for music recommendation by enriching the LastFM-1K dataset with semantic metadata, audio representations, lyric embeddings, and engagement signals. We evaluate these features across ItemID backbones and LLMs under zero-shot and fine-tuned settings. Our results show that multimodal enrichment generally improves recommendation quality, with the strongest gains when the ItemID baseline is weakest, such as SASRec with fine-tuned Qwen2.5-7B-Instruct, showing that semantic and acoustic information provide complementary signals beyond conventional interaction-only ItemID representations.

However, the observed improvements are not consistent across all backbone and encoder combinations, suggesting that current multimodal fusion approaches may introduce noisy or weakly aligned information. Future work includes exploring more robust modality alignment and fusion strategies, investigating larger-scale multilingual settings, and evaluating performance under cold-start recommendation scenarios.
\clearpage
\appendices

\section{Additional Implementation Details}
Our framework is implemented using PyTorch for model training and inference, HuggingFace Transformers and PEFT for LLM loading and LoRA-based fine-tuning, and NumPy and Pandas for data preprocessing and feature management. Handcrafted acoustic descriptors (MFCCs, spectral statistics, tempo, etc.) are computed with Librosa.

Pre-trained audio encoders (\textit{CLAP}, \textit{MERT}, \textit{Music2Vec}, and \textit{EnCodec}) and text encoders (\textit{MiniLM}, \textit{BGE-M3}, \textit{MPNet}, \textit{MultiLG}, and \textit{BERT}) are loaded from the HuggingFace Hub. Lyrics are retrieved through a three-stage fallback that queries the LRCLIB API, then the Genius API, and finally transcribes the audio with Whisper large-v3-turbo when neither API returns lyrics, keeping a source tag for each track.

The recommendation LLM backbones evaluated in this work are \textit{LLaMA-2-7B}, \textit{LLaMA-2-13B}, \textit{LLaMA-3-70B}, and \textit{Qwen2.5-7B-Instruct}. All four run zero-shot, and LLaMA-2-13B and Qwen2.5-7B-Instruct are also fine-tuned with LoRA. Larger backbones are loaded with 4-bit quantization where necessary to fit within available GPU memory. The MGPHot perceptual annotations used throughout the paper are produced with \textit{Azure OpenAI GPT-5} via the Batch API (Sec.~\ref{sec:LLM-gen-features}).

Audio retrieval and download pipelines use yt-dlp against the LastFM-1K track catalog. Structured audio attributes (e.g., valence, energy, danceability) are retrieved via the ReccoBeats and SpotifyEA APIs for each of the selected tracks.

\section{MGPHot Annotation Validation}
\label{app:mgphot-validation}

To assess the fidelity of our LLM-generated MGPHot annotations, we compare them against the published MGPHot ground-truth ratings~\cite{oramas2025mgphot} on 2,195 held-out tracks covered by both our top-50k LastFM-1K song list and the MGPHot release (inner join on the artist/title pair). For each of the 58 MGPHot attributes, we compute Spearman's rank correlation $\rho$ (rank-order agreement between predicted and ground-truth ratings) and mean absolute error (MAE) on the $[0,1]$-normalized scale, then aggregate by MGPHot category. Both metrics are evaluated using the prompt in Fig.~\ref{fig:mgphot-prompt}, which is the prompt that produced the annotations used throughout the paper.

Figure~\ref{fig:mgphot-eval-summary} visualizes the per-category results of Table~\ref{tab:MGPHot-validation} for the seven MGPHot categories of Table~\ref{tab:mgphot-cats}. Across every category the mean Spearman's $\rho$ is positive and the mean MAE stays well below $0.25$ on the $[0,1]$ scale, with an overall $\rho$ of 0.45 and MAE of 0.14, indicating that the LLM-generated annotations preserve the rank-order of the ground-truth ratings while remaining within a tight absolute-error band. Agreement is strongest for the lexically grounded categories, \textit{Lyrics} ($\rho = 0.58$) and \textit{Sonority} ($\rho = 0.54$), followed by \textit{Instrumentation} ($\rho = 0.52$). The lowest-agreement categories are the perceptually subtle and underrepresented ones, \textit{Rhythm} and \textit{Composition} (both $\rho = 0.31$) and \textit{Harmony} ($\rho = 0.40$, only $n=2$ attributes), consistent with the intuition that lexically grounded properties are easier for an LLM to infer from title-and-artist context than fine-grained acoustic structure.

\begin{table}[H]
\centering
\small
\renewcommand{\arraystretch}{0.9}
\setlength{\tabcolsep}{4pt}
\caption{LLM Ratings vs.\ MGPHot Ground Truth. Range is the Min and Max Per-Attribute $\rho$}
\label{tab:MGPHot-validation}
\begin{tabular}{lrrrr}
\toprule
\textbf{Category} &
\textbf{Attr.} &
\textbf{Avg. $\rho$} &
\textbf{Range} &
\textbf{MAE} \\
\midrule
Lyrics          & 8  & 0.58 & 0.45--0.73 & 0.15 \\
Sonority        & 6  & 0.54 & 0.15--0.82 & 0.12 \\
Instrumentation & 16 & 0.52 & 0.21--0.76 & 0.15 \\
Vocals          & 7  & 0.41 & 0.26--0.55 & 0.17 \\
Harmony         & 2  & 0.40 & 0.30--0.50 & 0.18 \\
Rhythm          & 10 & 0.31 & 0.11--0.57 & 0.09 \\
Composition     & 9  & 0.31 & 0.11--0.59 & 0.15 \\
\midrule
\textbf{All}    & \textbf{58} & \textbf{0.45} &
\textbf{0.11--0.82} & \textbf{0.14} \\
\bottomrule
\end{tabular}
\end{table}

\begin{figure}[H]
    \centering
    \includegraphics[width=\columnwidth]{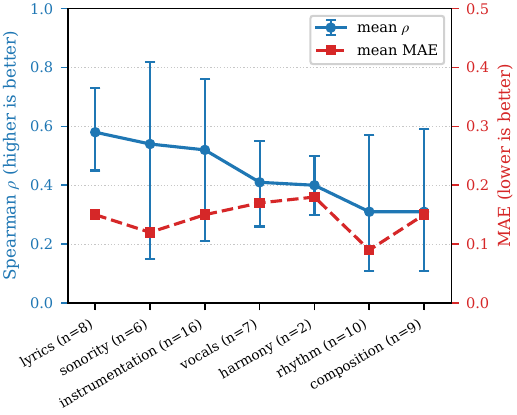}
    \caption{MGPHot validation, aggregated across the seven attribute categories of Table~\ref{tab:mgphot-cats}. Blue (left axis): mean Spearman's $\rho$ between LLM-generated and ground-truth ratings, higher is better. Red (right axis): mean MAE on $[0,1]$, lower is better. Error bars span the minimum and maximum per-attribute $\rho$ across the $n$ attributes in each category.}
    \label{fig:mgphot-eval-summary}
\end{figure}

\balance
\bibliographystyle{IEEEtranN}
\bibliography{main}

\end{document}